\documentclass[10pt,draftcls,onecolumn]{IEEEtran}
\usepackage{amsmath}
\usepackage{amssymb}
\usepackage{mathrsfs}
\usepackage{cite}
\usepackage{epsfig}
\usepackage{epsf}
\usepackage{theorem}
\usepackage{graphics}
\usepackage{subfigure}
\usepackage{url}
\usepackage{float}
\usepackage{multirow}
\usepackage{array}
\usepackage{hyperref}
\usepackage{color}
\definecolor{Blue}{rgb}{1,0,0}

\renewcommand{\baselinestretch}{1.0}

\newtheorem{lemma}{Lemma}
\newcommand{\vc}[1]{\boldsymbol{#1}}

\begin{document}
\title{On the Capacity of the Half-Duplex Diamond Channel\\
\let\thefootnote\relax\footnotetext{This work is financially supported by Nortel Networks and the corresponding matching funds by the Natural Sciences
and Engineering Research Council of Canada (NSERC), and Ontario Centers of Excellence (OCE).}}

\author{ Hossein Bagheri, Abolfazl
S. Motahari, and Amir K. Khandani \\
\small Coding and Signal Transmission Laboratory (www.cst.uwaterloo.ca)\\
Department of Electrical and Computer Engineering, University of Waterloo\\
Waterloo, Ontario, Canada, N2L 3G1 \\
Tel: 519-884-8552, Fax: 519-888-4338\\
Emails: \{hbagheri, abolfazl,
khandani\}@cst.uwaterloo.ca}

\date{}
\maketitle

\begin{abstract}
In this paper, a dual-hop communication system composed of a source
${\cal S}$ and a destination ${\cal D}$ connected through two
non-interfering half-duplex relays, ${\cal R}_1$ and ${\cal R}_2$,
is considered. In the literature of Information Theory, this
configuration is known as the \emph{diamond channel}. In this setup,
four \emph{transmission modes} are present, namely: 1) ${\cal S}$
transmits, and ${\cal R}_1$ and ${\cal R}_2$ listen (broadcast
mode), 2) ${\cal S}$ transmits, ${\cal R}_1$ listens, and
simultaneously, ${\cal R}_2$ transmits and ${\cal D}$ listens. 3)
${\cal S}$ transmits, ${\cal R}_2$ listens, and simultaneously,
${\cal R}_1$ transmits and ${\cal D}$ listens. 4) ${\cal R}_1$,
${\cal R}_2$ transmit, and ${\cal D}$ listens (multiple-access
mode). Assuming a constant power constraint for all transmitters, a
parameter $\Delta$ is defined, which captures some important
features of the channel. It is proven that for $\Delta\!=\!0$ the
capacity of the channel can be attained by successive relaying,
\emph{i.e,} using modes 2 and 3 defined above in a successive
manner. This strategy may have an infinite gap from the capacity of
the channel when $\Delta\!\neq\! 0$. To achieve rates as close as
0.71 bits to the capacity, it is shown that the cases of
$\Delta\!>\!0$ and $\Delta\!<\!0$ should be treated differently.
Using new upper bounds based on the dual problem of the linear
program associated with the cut-set bounds, it is proven that the
successive relaying strategy needs to be enhanced by an additional
broadcast mode (mode 1), or multiple access mode (mode 4), for the
cases of $\Delta\!<\!0$ and $\Delta\!>\!0$, respectively.
Furthermore, it is established that under average power constraints
the aforementioned strategies achieve rates as close as 3.6 bits to
the capacity of the channel.
\end{abstract}

\begin{center}
\end{center}

\section{Introduction}
\subsection{Motivation}
Relay-aided wireless systems, also called multi-hop systems, are
implemented to increase the coverage and the throughput of
communication systems \cite{PabstComMag04}. These systems are
becoming important parts of developing wireless communication
standards, such as IEEE 802.16j (also known as WiMAX)
\cite{802.16j}. Half-duplex relays, which transmit and receive data
in different times and/or frequencies, are proven to be more
practical and cost efficient in such standards than full-duplex
relays.

From information theoretical point of view, the capacity becomes
larger when more relays are added to the system. However, designing
optimum strategies, especially in half-duplex systems, is
challenging because subtle scheduling, \emph{i.e.}, timing among
transmission modes, is required to achieve rates near the capacity
of such systems. During the last decade, the main stream of research
carried out by several researchers dealt with single relay
communication systems (cf. \cite{IT05:Kramer} and references
therein). A simple model for investigating the potential benefits of
a system with multiple relays is a dual-hop configuration with two
parallel half-duplex relays (see Fig. 1). This configuration does
not cover all two-relay systems because there are no
source-destination and relay-relay links. However, it captures the
basic difficulty in finding the best strategy in the
system. As will be shown in this paper, a single strategy falls
short of achieving rates near the capacity of the system for all
channel realizations.

\subsection{History}
The single relay channel in which the relay facilitates a
point-to-point communication was first studied in
\cite{IT71:Vander}. Two important coding techniques,
\emph{decode-and-forward} and \emph{compress-and-forward}, were
proposed in \cite{IT79:Cover}. In the decode-and-forward scheme, the
relay decodes the received message. In the compress-and-forward
scheme, the relay sends the compressed (quantized) version of the
received data to the destination. Following \cite{IT79:Cover},
generalizations to multi-relay networks were investigated by several
researchers. A comprehensive survey of the progress in this area can
be found in \cite{IT05:Kramer}.

A simple model for understanding some aspects of the multi-relay
networks is a network with two parallel relays, as introduced in
\cite{ISIT00:Schein,PhD:Schein}, and Fig. 1. It is assumed that
there are no direct links from the source to the destination and
also between the relays. This channel is studied in
\cite{ISIT08:Kochman, Arxiv:Saeed09, PhD:Avestimehr, IT08:Kang,
Diamond:Xue, TWO-PATH:Rankov, Allerton07:Chang, Techrpt08:Saeed,
ISIT08:Ghabeli} and \cite{ISIT08:Avestimehr}, and referred to as the
\emph{diamond relay channel} in \cite{Diamond:Xue}.

For full-duplex relays, Schein and Gallager, in \cite{ISIT00:Schein}
and \cite{PhD:Schein}, provided upper and lower bounds on the
capacity of the diamond channel. In particular, they considered the
\emph{amplify-and-forward}, and the decode-and-forward schemes, as
well as a hybrid of them based on the time-sharing principle.
Kochman, \emph{et al.} proposed a \emph{rematch-and-forward} scheme
when different fractions of bandwidth can be allotted to the first
and second hops \cite{ISIT08:Kochman}. Rezaei, \emph{et al.}
suggested a \emph{combined amplify-and-decode-forward} strategy and
proved that their scheme always performs better than the
rematch-and-forward scheme \cite{Arxiv:Saeed09}. In addition, they
showed that the time-sharing between the combined
amplify-and-decode-forward and decode-and-forward schemes provides a
better achievable rate when compared to the time-sharing between the
amplify-and-forward and decode-and-forward, and also between the
rematch-and-forward and decode-and-forward, considered in
\cite{PhD:Schein}, and \cite{ISIT08:Kochman}, respectively. Kang and
Ulukus employed a combination of the decode-and-forward and
compress-and-forward schemes to obtain the capacity of a special
class of the diamond channel with a noiseless relay
\cite{IT08:Kang}. Ghabeli and Aref in \cite{ISIT08:Ghabeli} proposed
a new achievable rate based on the generalized block Markov encoding
\cite{IT82:Aref}. They also showed that their scheme achieves the
capacity of a class of deterministic relay networks.

Half-duplex relays are studied in \cite{PhD:Avestimehr,
Diamond:Xue,TWO-PATH:Rankov,Allerton07:Chang,Techrpt08:Saeed,
IT08:Gharan, IT08:Sreeram}. Xue and Sandhu in \cite{Diamond:Xue}
proposed several schemes including the multi-hop with spatial reuse,
scale-forward, broadcast-multiaccess with common message,
compress-and-forward, and hybrid methods. These authors demonstrated
that the \emph{multi-hop with spatial reuse} protocol can achieve
the channel capacity if the parallel links have the same capacity.
Unlike \cite{ISIT00:Schein, PhD:Schein, ISIT08:Kochman,
Arxiv:Saeed09, ISIT08:Avestimehr, IT08:Kang, Diamond:Xue}, which
assumed no direct link exists between the relays,
\cite{TWO-PATH:Rankov,Allerton07:Chang, Techrpt08:Saeed} considered
such link. More specifically, Chang, \emph{et al.} proposed a
combined dirty paper coding and block Markov encoding scheme
\cite{Allerton07:Chang}. Using numerical examples, they showed that
the gap between their proposed strategy and the upper bound is
relatively small in most cases. Rezaei, \emph{et al.} considered two
scheduling algorithms, namely \emph{successive} and
\emph{simultaneous} relaying \cite{Techrpt08:Saeed}. They derived
asymptotic capacity results for the successive relaying
 and also proposed an achievable rate for the simultaneous
relaying using a combination of the amplify-and-forward and
decode-and-forward schemes. Other related papers are
\cite{IT08:Gharan, IT08:Sreeram, ISIT02:Gastpar, WC07:Fan}.

Characterizing the capacity of an information theoretic channel may
be difficult. A simpler, yet important approach is to find an
achievable scheme that ensures a small gap from the capacity of the
channel. Recently, Etkin \emph{et al.} characterized the capacity
region of the interference channel to within one bit
\cite{EtkinIT08}. Following this new capacity analysis perspective,
Avestimehr \emph{et al.} proposed a deterministic model to better
analyze the general single-source single-destination and the
single-source multi-destination Gaussian networks
\cite{ISIT08:Avestimehr, PhD:Avestimehr}. Their
\emph{quantize-and-map} achievablity scheme is guaranteed to provide
a rate that is within a constant number of bits (determined by the
graph topology of the network) from the cut-set upper bound.
\begin{figure}
\centering
\includegraphics[width=0.5\textwidth]{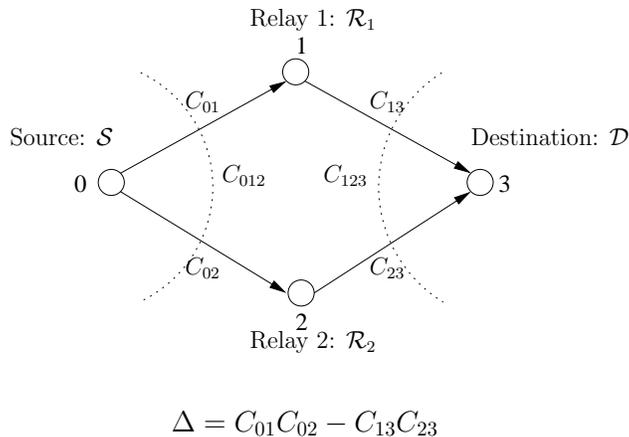}
\label{fig: model}
\caption{The diamond channel with its fundamental parameter $\Delta$.}
\end{figure}

\subsection{Relation to Previous Works}
In this paper, the setup and assumptions used in \cite{Diamond:Xue},
with no link between the relays, are followed. In
\cite{Diamond:Xue}, the multi-hop with spatial reuse scheme proved
to achieve the capacity of the diamond channel if the capacities of
the parallel links in Fig. 1 are equal. This is called the
\emph{Multi-hopping Decode-and-Forward} (MDF) scheme. In the MDF
scheme, relays successively forward their decoded messages to the
destination (see Forward Modes I and II in Fig. 2). By introducing a
fundamental parameter of the channel $\Delta$ (see Fig. 1), we
generalize the optimality condition of the MDF scheme. In
particular, we show that whenever $\Delta=0$, the cut-set upper bound
can be achieved. We also show that the MDF scheme cannot have a
small gap from the cut-set upper bound for all channel realizations
because the optimum strategy is highly related to the value of
$\Delta$.

In \cite{PhD:Avestimehr}, the aim has been to establish the constant
gap argument for the general relay networks with a single source and
not to obtain a small gap optimized for a specific channel, such as
the diamond channel. For the half-duplex diamond channel, the
expression for the gap derived in \cite{PhD:Avestimehr} results in a
6-bit gap. In this paper, however, we focus on the diamond channel
and obtain a smaller gap using our proposed achievablity scheme. In
addition, we provide closed-form expressions for the time intervals
associated with the transmission modes in the proposed scheduling.
Specifically, we show that the expressions are different from those
of the cut-set upper bound. This is in contrast to
\cite{PhD:Avestimehr}, where the constant gap between the cut-set
bound and the quantize-and-map scheme was assured for every fixed
scheduling, including the optimum scheduling associated with the
cut-set upper bound.

In \cite{PhD:Avestimehr}, using a different achievablity (a partial
decode-and-forward) scheme than the quantize-and-map scheme,
Avestimehr \emph{et al.} showed that the capacity of the
\emph{full-duplex} diamond channel can be characterized within 1 bit
per real dimension, regardless of the values of the channel gains.
However, applying this scheme to the \emph{half-duplex} diamond
channel does not guarantee a constant gap from the channel capacity.
We take one further step by providing an achievable scheme that
ensures a small gap from the upper bounds for the half-duplex
diamond channel. In particular, we show that the gap is smaller than
.71 bits, assuming all transmitters have constant power constraints.
We also prove that when transmitters have average power constraints
instead, the gap is less than 3.6 bits.

The rest of this paper is organized as follows: Section II
introduces the system model, the main ideas and results of this
work. Section III presents the MDF scheme, which achieves the
channel capacity for $\Delta=0$. Sections IV and V provide the
achievable schemes, upper bounds, and gap analysis for $\Delta<0$
and $\Delta>0$ cases, respectively. Section VI concludes the paper.
In addition, Appendix \ref{appendix: GDOF} characterizes the
Generalized Degrees Of Freedom (GDOF) of the diamond channel to
obtain asymptotic capacity of the channel. Finally, Appendix
\ref{appendix: general average power} addresses the diamond channel
with average power constraints.

\subsection{Notations}
Throughout the paper, $\bar{x}\!\triangleq\!1\!-\!x$, and $x^{*}$
denotes the optimal solution to an optimization problem with an
objective function $F(x)$. The transpose of the vector or matrix
$\textbf{A}$ is indicated by $\textbf{A}^{T}$. $a\!\Rightarrow\!b$
represents the link from node $a$ to node $b$. Also,
$x\!\leftrightarrow\!y$ means that the roles of $x$ and $y$ are
exchanged in a given function $F(x,y)$. In addition, it is assumed
that all logarithms are to base $2$. Finally,
$\mathcal{C}(P)\!\triangleq\!\frac{1}{2}\log\left(1+P\right)$.

\section{Problem Statement and Main Results}
In this work, a dual-hop communication system, depicted in Fig. 1,
is considered. The model consists of a source (${\cal S}$), two
parallel half-duplex relays (${\cal R}_1$, ${\cal R}_2$), and a
destination (${\cal D}$), respectively, indexed by 0, 1, 2, and 3 as
shown in Fig. 1. No link is assumed between Source and Destination,
 as well as between the relays. The channel gain between
node $a$ and $b$ is assumed to be constant, known to all nodes, and
is represented by $h_{ab}$ with magnitude $\sqrt{g_{ab}}$.

Due to the half-duplex constraint, four transmission modes exist in
the diamond channel where, in every mode, each relay either
transmits data to Destination or receives data from Source
(see Fig. 2). In the figure, $X_{a}^{(i)}$ and $Y_{a}^{(i)}$
represent the transmitting and receiving signals at node $a$
corresponding to mode $i$, respectively. The total transmission time
is normalized to one and partitioned into four time intervals
($t_1,t_2,t_3,t_4$) corresponding to modes $1, 2, 3$, and $4$, with
the constraint $\sum_{i=1}^{4}
t_{i}\!=\!1$.
 \begin{figure}
\centering
\includegraphics[width=1.0\textwidth]{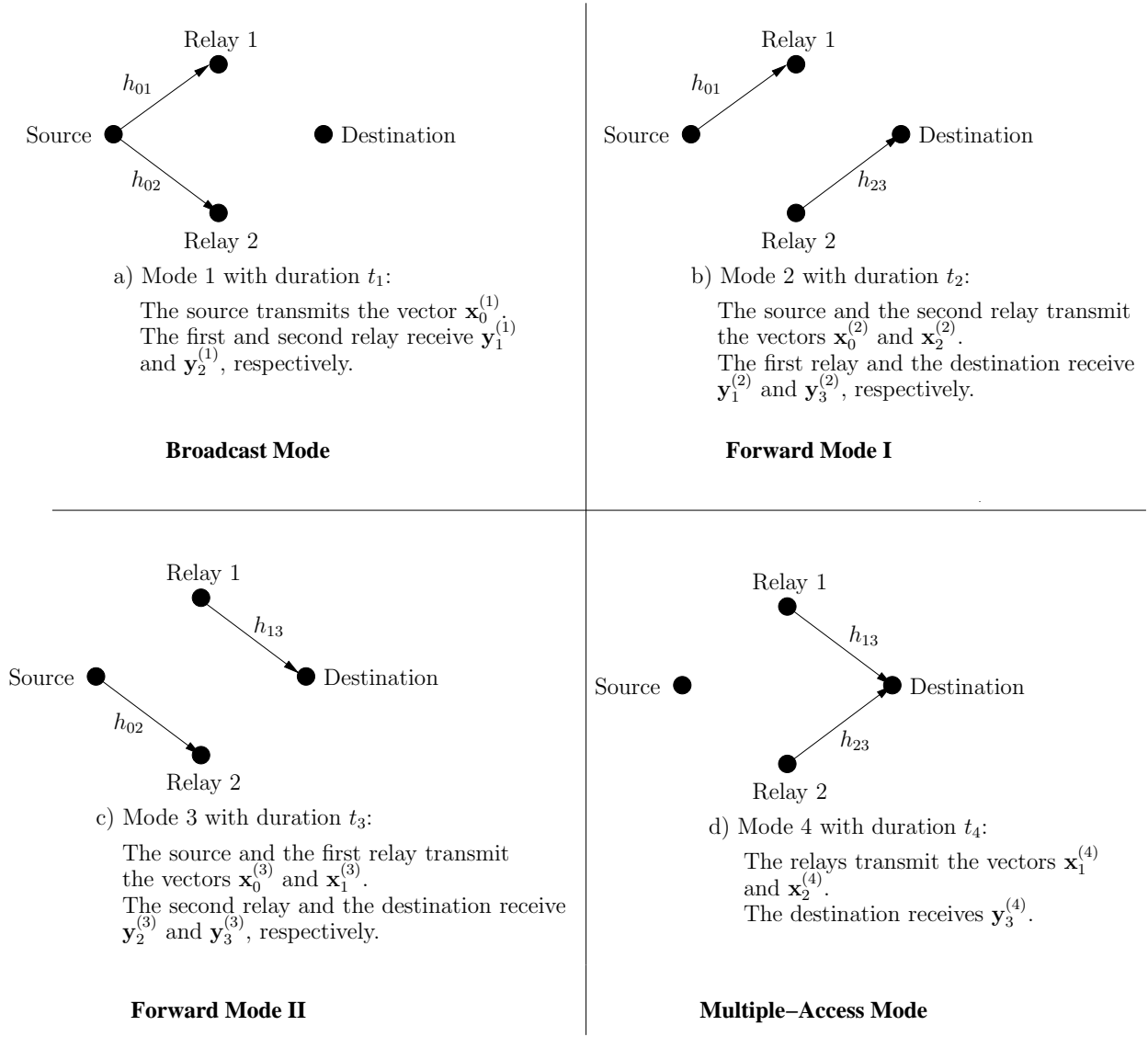}
\caption{Transmission modes for the diamond channel.}
\end{figure}\label{fig: states}
The discrete-time baseband representation of the received signals at
Relay 1, Relay 2, and Destination are respectively given by:
\begin{IEEEeqnarray*}{rl}
\label{eq01a} Y_{1} & =  h_{01}  X_{0} + N_{1}, \nonumber\\
\label{eq02a} Y_{2} & =  h_{02}  X_{0} +N_{2},\nonumber\\
\label{eq03a} Y_{3} & =  h_{13}  X_{1} + h_{23}  X_{2} +N_{3},
\end{IEEEeqnarray*}
where $N_{a}$ is the Gaussian noise
at node $a$ with unit variance.

Let us assume Source, Relay 1, and Relay 2 consume, respectively,
$P_{{\cal S}}^{(i)}$, $P_{{\cal R}_{1}}^{(i)}$, and $P_{{\cal
R}_{2}}^{(i)}$ amount of power in mode $i$, \emph{i.e.,}
\begin{IEEEeqnarray*}{rl}
  \frac{1}{t_{i}}\sum_{t_{i}} \mid\! X_{0} \! \mid^{2} &\leq P_{{\cal S}}^{(i)}, \nonumber \\
  \frac{1}{t_{i}}\sum_{t_{i}} \mid\! X_{1} \! \mid^{2} &\leq P_{{\cal R}_1}^{(i)}, \nonumber \\
    \frac{1}{t_{i}}\sum_{t_{i}} \mid\! X_{2} \! \mid^{2} &\leq P_{{\cal R}_2}^{(i)}.
  \end{IEEEeqnarray*}

The total power constraints for Source, Relay 1, and Relay 2 are
$P_{{\cal S}}$, $P_{{\cal R}_1}$, and $P_{{\cal R}_2}$,
respectively, and are related to the amount of power spent in each
mode as follows:
\begin{IEEEeqnarray*}{rl}
  \sum_{i=1}^{4}t_{i} P_{{\cal S}}^{(i)} &\leq P_{{\cal S}}, \nonumber \\
  \sum_{i=1}^{4}t_{i} P_{{\cal R}_1}^{(i)} &\leq P_{{\cal R}_1}, \nonumber \\
  \sum_{i=1}^{4}t_{i} P_{{\cal R}_2}^{(i)} &\leq P_{{\cal R}_2}.
  \end{IEEEeqnarray*}
Due to some practical considerations on the power constraints
\cite{Diamond:Xue}, we mainly consider \emph{constant} power
constraints for transmitters, \emph{i.e.}, for $i \in
\{1,\cdots,4\}$,
  \begin{IEEEeqnarray}{rl}
  P_{{\cal S}}^{(i)} &= P_{{\cal S}},  \nonumber\\
  \label{eq: constant_power} P_{{\cal R}_1}^{(i)} &= P_{{\cal R}_1}, \\
  P_{{\cal R}_2}^{(i)} &= P_{{\cal R}_2}. \nonumber
  \end{IEEEeqnarray}
Without loss of generality, a unit power constraint is considered for all nodes, \emph{i.e.}, $P_{{\cal S}}=P_{{\cal R}_1}=P_{{\cal R}_2}=1$. We define the parameters $C_{01},C_{02},C_{13},C_{23}$ as $\mathcal{C}(g_{01}),\mathcal{C}(g_{02}),\mathcal{C}(g_{13}),\mathcal{C}(g_{23})$, respectively. Moreover, $C_{012}$ and $C_{123}$ are defined as:
 \begin{equation}
\begin{array}{rl}
  \label{eq: C012}C_{012}&\triangleq\mathcal{C}(g_{01}+g_{02}),\\
  C_{123}
  &\triangleq\mathcal{C}\big((\sqrt{g_{13}}+\sqrt{g_{23}})^{2}\big).
  \end{array}
  \end{equation}
The case in which transmitters have average power constraints
instead of constant power constraints is addressed in Appendix
\ref{appendix: general average power}.

In this work, we are interested in finding communication protocols
that operate close to the channel capacity. We introduce an important parameter of the channel $\Delta$ as:
\begin{equation}\label{eq: Delta}
    \Delta\triangleq C_{01}C_{02}-C_{13}C_{23}.
\end{equation}
We categorize all realizations of the diamond channel into three
groups based on the sign of $\Delta$ (\emph{i.e.}, $\Delta\!<\!0$,
$\Delta\!=\!0$, and $\Delta\!>\!0$). As will be shown in the sequel,
the sign of $\Delta$ plays an important role in designing the
optimum scheduling for the channel.

In this setup, the cut-set bounds can be stated in the form of a
Linear Program (LP) due to the assumption of constant power
constraints for all transmitters. By analyzing the dual program we
provide fairly tight upper bounds expressed as single equations
corresponding to different channel conditions. Using the dual
problem, we prove that when $\Delta\!=\!0$, the MDF scheme achieves
the capacity of the diamond channel. Note that $\Delta\!=\!0$
(\emph{i.e.}, $C_{01}C_{02}=C_{13}C_{23}$) includes the previous
optimality condition presented in \cite{Diamond:Xue} (\emph{i.e.},
$C_{01}\!=\!C_{23}$ and $C_{02}\!=\!C_{13}$) as a special case. To
realize how close the MDF scheme performs to the capacity of the
channel when $\Delta\!\neq\!0$, we calculate the gap from the upper
bounds. We show that the MDF scheme provides the gap of less than
$1.21$ bits when applied in the symmetric or some classes of
asymmetric diamond channels. More importantly, we explain that the
gap can be arbitrarily large for certain ranges of parameters.

By employing new scheduling algorithms we shrink the gap to $.71$
bits for \emph{all} channel conditions. In particular, for
$\Delta\!<\!0$ we add Broadcast (BC) Mode (shown in Fig. 2) to the
MDF scheme to provide the relays with more reception time. In this
three-mode scheme, referred to as Multi-hopping Decode-and-Forward
with Broadcast (\emph{MDF-BC}) scheme, the relays decode what they
have received from Source and forward the re-encoded information
to Destination in Forward Modes I and II. When $\Delta\!>\!0$,
Multiple-Access (MAC) mode (shown in Fig. 2) in which the relays transmit
independent information to Destination is added to the MDF scheme. We call this protocol
Multi-hopping Decode-and-Forward with Multiple-Access
(\emph{MDF-MAC}) scheme.

The mentioned contributions are associated with the case wherein the
transmitters are operating under constant power constraints
(\ref{eq: constant_power}). However, for a more general setting in
which the transmitters are subject to average power constraints
(\ref{eq: average_power}), it is shown in Appendix \ref{appendix:
general average power} that the cut-set upper bounds are increased
by at most $2.89$ bits. Therefore, the proposed achievable schemes
guarantee the maximum gap of 3.6 bits from the cut-set upper bounds
in the general average power constraint setting.

\subsection{Coding Scheme}
The proposed achievable scheme may employ all four transmission
modes as follows:
\begin{enumerate}
  \item \emph{\textbf{Broadcast Mode}}: In $t_{1}$ fraction of the transmission time, Source broadcasts
  independent information to Relays 1 and 2 using the superposition coding technique.
  \item \emph{\textbf{Forward Mode I}}: In $t_{2}$ fraction of the transmission time, Source transmits
  new information to Relay 1. At the same time, Relay 2 sends the re-encoded version of part of the data received during Broadcast Mode and/or Forward Mode II of the previous block to Destination.
  \item \emph{\textbf{Forward Mode II}}: In $t_{3}$ fraction of the transmission time, Source transmits new
  information to Relay 2. At the same time, Relay 1 sends the re-encoded version of part of what it has received during Broadcast Mode and/or Forward Mode I of the previous block to Destination.
  \item \emph{\textbf{Multiple-Access Mode}}: In the remaining $t_{4}$ fraction of the transmission time,
  Relays 1 and 2 simultaneously transmit the residual information (corresponding to the previous block) to
  Destination where, \emph{joint decoding} is performed to decode the received data.
\end{enumerate}

In Broadcast Mode, superposition coding, which is known to be the
optimal transmission scheme for the degraded broadcast channel
\cite{book:Cover}, is used to transmit independent data to the
relays. The resulting data-rates $u$ and $v$, respectively
associated with Relay 1 and Relay 2 are:
\begin{equation}
\label{eq: u_alpha}u(\eta)=\left\{ \begin{array}{ll}
\mathcal{C}(\eta g_{01}) & \textrm{if} \quad g_{02}\leq g_{01}\\
C_{01}-\mathcal{C}(\eta g_{01}) & \textrm{if} \quad g_{01}<g_{02},
\end{array} \right.
\end{equation}
\begin{equation}
\label{eq: v_alpha}v(\eta)=\left\{ \begin{array}{ll}
C_{02}-\mathcal{C}(\eta g_{02}) & \textrm{if} \quad
g_{02}\leq g_{01}\\
\mathcal{C}(\eta g_{02}) & \textrm{if} \quad g_{01}< g_{02}.
\end{array} \right.
\end{equation}
The power allocation parameter $\eta$ determines the amount of
Source power used to transmit information
 to the relay with better channel quality in Broadcast Mode.

In Multiple-Access Mode, a multiple-access channel exists in which
the users (relays) have independent messages for Destination. For
this channel, joint decoding is optimum, which provides the
following rate region \cite{book:Cover}:
\begin{equation}\label{eq: MAC-Cap-Reg}
\begin{array}{rl}
R_{1}&\leq t_{4}C_{13},\\
R_{2}&\leq t_{4}C_{23},\\
R_{1}+R_{2}&\leq t_{4}C_{\text{MAC}},
\end{array}
\end{equation}
where $R_{1}, R_{2}$ are the rates that Relay 1 and Relay 2 provide
to Destination in Multiple-Access Mode, respectively, and
$C_{\text{MAC}}$ is defined as:
\begin{equation}\label{eq: C_MAC}
    C_{\text{MAC}}\triangleq \mathcal{C}(g_{13}+g_{23}).
\end{equation}

According to the protocol, Relay 1 can receive up to
$t_{1}u+t_{2}C_{01}$ bits per channel use during Broadcast Mode and
Forward Mode I. Then the relay has the opportunity to send its
received information to Destination in Forward Mode II and
Multiple-Access Mode, with the rate $t_{3}C_{13}+R_{1}$. Similarly,
Relay 2 can receive and forward messages with the rates
$t_{1}v+t_{3}C_{02}$, and $t_{2}C_{23}+R_{2}$, respectively.
Therefore, the maximum achievable rate of the scheme, $R$, is:
\begin{equation}\label{eq: Ach_Gen}
    R=\max_{\scriptstyle{\sum_{i=1}^{4}t_{i}= 1, t_{i}\geq 0}
    }\!\bigg\{\!\min\{t_{1}u+t_{2}C_{01}, t_{3}C_{13}+R_{1}\}+ \min\{t_{1}v+t_{3}C_{02},
    t_{2}C_{23}+R_{2}\!\}\bigg\}.
\end{equation}

Sections \ref{sec: Delta0}-\ref{sec: MDF-MAC} show that employing
Forward Modes I and II for $\Delta\!=\!0$, the first three
transmission modes for $\Delta\!<\!0$, and the last three
transmission modes for $\Delta\!>\!0$ are sufficient to achieve a
small gap from the derived upper bounds.
\subsection{Cut-set Upper Bound and the Dual Program}\label{subsec: Cut-set Upper Bound}
For general half-duplex networks with $K$ relays, Khojastepour
\emph{et al.} proposed a cut-set type upper bound by doing the
following steps:
 \begin{enumerate}
 \item Fix the input distribution and scheduling, \emph{i.e.}, $p(X_0,X_1,X_2)$, and $t_1,t_2, t_3, t_4$ such that $\sum_{i=1}^{4}t_{i}= 1$.
   \item Find the rate $R_{i,j}$ associated with the cut $j$ for each transmission mode $i$ where $i, j
   \in\{1,\cdots,2^{K}\}$.
   \item Multiply $R_{i,j}$ by the corresponding time interval
   $t_{i}$.
   \item Compute $\sum_{i=1}^{2^{K}} t_{i}R_{i,j}$ and minimize it over all cuts.
   \item Take the supremum over all input distributions and schedulings.
 \end{enumerate}
The preceding procedure can be directly applied to the diamond
channel, whose transmission modes are shown in Fig. 2. The best
input distribution and scheduling lead to:
\begin{equation}
\begin{array}{rl}
&C_{\text{DC}} \leq
t_{1}I(X_{0}^{(1)};Y_{1}^{(1)},Y_{2}^{(1)})+t_{2}I
(X_{0}^{(2)};Y_{1}^{(2)}|X_{2}^{(2)})+t_{3}I(X_{0}^{(3)};Y_{2}^{(3)}|X_{1}^{(3)})+t_{4}.0,\nonumber\\
 &C_{\text{DC}} \leq
t_{1}I(X_{0}^{(1)};Y_{1}^{(1)})+t_{2}\Big(I (X_{0}^{(2)};Y_{1}^{(2)})+I (X_{2}^{(2)};Y_{3}^{(2)})\Big)+t_{3}.0+t_{4}I (X_{2}^{(4)};Y_{3}^{(4)}|X_{1}^{(4)}),\nonumber\\
&C_{\text{DC}} \leq
t_{1}I(X_{0}^{(1)};Y_{2}^{(1)})+t_{2}.0+t_{3}\Big(I
(X_{0}^{(3)};Y_{2}^{(3)})+I (X_{1}^{(3)};Y_{3}^{(3)})\Big)+t_{4}I
(X_{1}^{(4)};Y_{3}^{(4)}|X_{2}^{(4)}),\nonumber\\
 &C_{\text{DC}} \leq
t_{1}.0+t_{2}I(X_{2}^{(2)};Y_{3}^{(2)})+t_{3}I
(X_{1}^{(3)};Y_{3}^{(3)})+t_{4}I
(X_{1}^{(4)},X_{2}^{(4)};Y_{3}^{(4)}),
 \end{array}
  \end{equation}
where $C_{DC}$ denotes the capacity of the diamond channel. The above bounds do not decrease if each mutual information term is replaced by its maximum value. This substitution simplifies the
computation of the upper bound, called $R_{\text{up}}$, by providing the following LP \cite{Diamond:Xue}:
\begin{equation}\label{eq: cut-set-org}
\begin{array}{ll}
\text{maximize} &R_{\text{up}}  \\
\text{subject to:}~&R_{\text{up}} \leq t_{1}C_{012}+t_{2}C_{01}+t_{3}C_{02}+t_{4}.0 \\
&R_{\text{up}} \leq t_{1}C_{01}+t_{2}(C_{01}+C_{23})+t_{3}.0+t_{4}C_{23} \\
&R_{\text{up}} \leq t_{1}C_{02}+t_{2}.0+t_{3}(C_{02}+C_{13})+t_{4}C_{13} \\
&R_{\text{up}} \leq t_{1}.0+t_{2}C_{23}+t_{3}C_{13}+t_{4}C_{123} \\
&\sum_{i=1}^{4}t_{i}= 1,~t_{i}\geq 0.
\end{array}
\end{equation}
To obtain appropriate single-equation upper bounds on the capacity,
we rely on the fact that every feasible point in the dual program
provides an upper bound on the primal. Hence, we develop the desired
upper bounds by looking at the dual program. In the sequel, we
derive the dual program for the LP (\ref{eq: cut-set-org}).

We start with writing the LP in the standard form as:
\begin{equation*}
\begin{array}{ll}
\text{maximize} &\textbf{c}^{T}\textbf{x}  \\
\text{subject to:}~&\textbf{A}\textbf{x}\leq \textbf{b}\\
&\textbf{x}\geq 0,
\end{array}
\end{equation*}
where the unknown vector $\textbf{x}\!=\![t_{1}, t_{2}, t_{3}, t_{4},
R_{\text{up}}]^{T}$, the vectors of coefficients
$\textbf{b}\!=\!\textbf{c}\!=\![0, 0, 0, 0, 1]^{T}$, and the matrix of
coefficients $\textbf{A}$ is:
\begin{IEEEeqnarray*}{ll}
\textbf{A}&=\left(
  \begin{array}{ccccc}
    -C_{012} & -C_{01} & -C_{02} & 0 & 1\\
    -C_{01} & -(C_{01}+C_{23}) & 0 & -C_{23} & 1 \\
    -C_{02} & 0 & -(C_{02}+C_{13}) & -C_{13} & 1 \\
    0 & -C_{23} & -C_{13} & -C_{123} & 1 \\
    1 & 1 & 1 & 1 & 0 \\
  \end{array}
\right).
\end{IEEEeqnarray*}
Since $\textbf{A}=\textbf{A}^{T}$, it is easy to verify that the
primal and dual programs share the same
form, \emph{i.e.},
\begin{equation}\label{eq: dual}
\begin{array}{ll}
\text{minimize} &R_{\text{up}} \\
\text{subject to:}~& R_{\text{up}} \geq \tau_{1}C_{012}+\tau_{2}C_{01}+\tau_{3}C_{02}+\tau_{4}.0 \\
&R_{\text{up}} \geq \tau_{1}C_{01}+\tau_{2}(C_{01}+C_{23})+\tau_{3}.0+\tau_{4}C_{23} \\
&R_{\text{up}} \geq \tau_{1}C_{02}+\tau_{2}.0+\tau_{3}(C_{02}+C_{13})+\tau_{4}C_{13} \\
&R_{\text{up}} \geq \tau_{1}.0+\tau_{2}C_{23}+\tau_{3}C_{13}+\tau_{4}C_{123} \\
&\sum_{i=1}^{4}\tau_{i}= 1,~\tau_{i}\geq 0.
\end{array}
\end{equation}
In the dual program (\ref{eq: dual}), $\tau_{i}$, for
$i\!\in\!\{1,\cdots,4\}$ corresponds to the $i$th rate constraint in
the primal LP (\ref{eq: cut-set-org}). Clearly, the LP (\ref{eq:
cut-set-org}) is \emph{feasible}. Hence, the duality of linear
programming ensures that there is no gap between the primal and the
dual solutions \cite{book:LP}. However, the benefit of using the
dual problem here is that any feasible choice of the vector
$\vc{\tau}$ provides an upper bound to the rate obtained by solving
the original LP. This property is known as the weak duality property
of LP \cite{book:LP}. Appropriate vectors (\emph{i.e.},
$\vc{\tau}$'s) in the dual program (\ref{eq: dual}) are selected to
obtain fairly tight upper bounds. In fact, employing such vectors
instead of solving the primal LP (\ref{eq: cut-set-org}) simplifies
the gap analysis. In sections \ref{sec: MDF-BC} and \ref{sec:
MDF-MAC}, these vectors are provided for $\Delta\!<\!0$ and
$\Delta\!>\!0$ cases, respectively. In the following sections, we
employ the proposed achievable schemes together with the derived
upper bounds to characterize the capacity of the diamond channel up
to 0.71 bits.
\section{MDF Scheme and Achieving the Capacity for $\Delta=0$}\label{sec: Delta0}
In this section, the MDF scheme is described and then proved to be
capacity-achieving when $\Delta\!=\!0$.
\subsection{MDF Scheme}\label{subsection: MDF-Ach}
The MDF scheduling algorithm uses two transmission modes: Forward
Modes I and II shown in Fig. 2 along with the decode-and-forward
strategy and can be described as follows:
\begin{enumerate}
  \item In $\lambda$ fraction of the transmission time, Source and Relay 2
  transmit to Relay 1 and Destination, respectively.
  \item In the remaining $\bar{\lambda}$ fraction of the transmission time, Source and Relay 1 transmit to Relay 2 and Destination, respectively.
\end{enumerate}

The achievable rate of the MDF scheme is the summation of the rates
of the first and second parallel paths (branches) from Source to
Destination, which can be expressed as\cite{Diamond:Xue}:
\begin{equation*}
    R_{\textrm{MDF}}=\max_{0\leq\lambda\leq1}\Big\{\min\{\lambda C_{01},\bar{\lambda} C_{13}\}+\min\{\bar{\lambda} C_{02},\lambda C_{23}\}\Big\}.
\end{equation*}
The above LP can be re-written as:
\begin{equation*}
\begin{array}{ll}
\text{maximize} &R_{1}+R_{2}  \\
\text{subject to:}~&R_{1} \leq \lambda C_{01} \\
&R_{1} \leq \bar{\lambda} C_{13}\\
&R_{2} \leq \bar{\lambda} C_{02}\\
&R_{2} \leq \lambda C_{23}\\
&0 \leq \lambda\leq 1,
\end{array}
\end{equation*}
where $R_{1}$ and $R_{2}$ denote the rate of the upper and the lower
branches, respectively. This LP has three unknowns
($R_{1},R_{2},\lambda$) and six inequalities. The solution turns
three out of six inequalities to equality. The optimum time interval
$\lambda^{*}$ can not be equal to 0 or 1, as both solutions give a
zero rate. Hence, three out of the first four inequalities should
become equality, which leads to the following achievable rates for
different channel conditions:
\begin{align}\label{eq: MDF_rate}
R_{\text{MDF}} = \left\{ \begin{array}{ll}
R_{\text{MDF}}^{1}=\frac{\textstyle{C_{01}(C_{02}+C_{13})}}{\textstyle{C_{01}+C_{13}}} & \textrm{if $\Delta\!\leq\!0$,\ $C_{02}\!\leq\!C_{01}$ }\vspace*{6pt}\\
R_{\text{MDF}}^{2}=\frac{\textstyle{C_{02}(C_{01}+C_{23})}}{\textstyle{C_{02}+C_{23}}} & \textrm{if $\Delta\!\leq\!0$,\ $C_{02}\!>\!C_{01}$ }\vspace*{6pt}\\
R_{\text{MDF}}^{3}=\frac{\textstyle{C_{13}(C_{01}+C_{23})}}{\textstyle{C_{01}+C_{13}}} & \textrm{if $\Delta\!>\!0$,\ $C_{23}\!\leq\!C_{13}$ }\vspace*{6pt}\\
R_{\text{MDF}}^{4}=\frac{\textstyle{C_{23}(C_{02}+C_{13})}}{\textstyle{C_{02}+C_{23}}}
& \textrm{if $\Delta\!>\!0$,\ $C_{23}\!>\!C_{13}$. }
\end{array} \right.
\end{align}
In particular, the achievable rate for the \emph{symmetric} diamond
channel, in which $C_{01}=C_{02}$ and $C_{13}=C_{23}$, is:
\begin{equation*}
    R_{\text{MDF}}^{\text{sym}}=\min\{C_{01},C_{13}\}.
\end{equation*}

The optimum time interval $\lambda^{*}$ is either equal to $\lambda_{1}^{*}$ or $\lambda_{2}^{*}$ defined below:
\begin{IEEEeqnarray*}{rl}
\lambda^{*}&=\left\{
              \begin{array}{ll}
                \frac{C_{13}}{C_{01}+C_{13}}\triangleq \lambda_{1}^{*}, \\
                \quad \ \ \text{or} \\
                \frac{C_{02}}{C_{02}+C_{23}}\triangleq \lambda_{2}^{*}.
              \end{array}
            \right.
\end{IEEEeqnarray*}
Note that if $\lambda^{*}\!=\!\lambda_{1}^{*}$, then $\lambda_{1}^{*}C_{01}\!=\!\bar{\lambda}_{1}^{*}C_{13}$. Similarly, $\lambda^{*}\!=\!\lambda_{2}^{*}$ leads to $\bar{\lambda}_{2}^{*}C_{02}\!=\!\lambda_{2}^{*}C_{23}$. In other words, $\lambda_{i}^{*}$ for $i\in\{1,2\}$ makes the maximum amount of data that can be received by Relay $i$ equal to the maximum amount of data that can be forwarded by Relay $i$. In this case, branch $i$ (composed of $0\!\Rightarrow\!i\!\Rightarrow\!3$ links) is said to be \emph{fully utilized}.

It is interesting to consider that the case fully utilizing branch 1
or branch 2 leads to the same data-rate. This case occurs when one
of the following happens:
\begin{equation}\label{eq: rate_eq_2}
\left\{
\begin{array}{rl}
\Delta&=0,\\
C_{01}&=C_{02} \quad \textrm{if} \ \Delta<0 ,\\
C_{13}&=C_{23} \quad \textrm{if} \ \Delta>0.
\end{array}
\right.
\end{equation}
In these situations, one can use either $\lambda_{1}^{*}$ or
$\lambda_{2}^{*}$ fraction of the transmission time for Forward Mode
I and the remaining fraction for Forward Mode II and achieve the
same data-rate. It will be shown later that the MDF scheme achieves
the capacity of the diamond channel if $\Delta\!=\!0$ and is at most
1.21 bits less than the capacity for the other two cases. It is
remarked that $\Delta\!=\!0$ makes both branches fully utilized and
all four rates in Eq. (\ref{eq:
MDF_rate}) equal.

\subsection{MDF is Optimal for $\Delta\!=\!0$}
Here, it is explained that $R_{\text{up}}^*$, found by solving the
dual-program (\ref{eq: dual}), is the same as the MDF rate given in
Eq. (\ref{eq: MDF_rate}) for $\Delta\!=\!0$. It is easy to observe
that
\begin{equation}\label{eq: cut-set-opt-timevec}
    \vc{\tau}^{*}=\left[0,\frac{C_{13}}{C_{01}+C_{13}},\frac{C_{23}}{C_{02}+C_{23}},0\right]
\end{equation}
makes all four rate constraints in the dual-program (\ref{eq: dual})
equal to the rate obtained in Eq. (\ref{eq: MDF_rate}) and satisfies
$\sum_{i=1}^{4}\tau_{i}\!=\!1$. Therefore, the upper bound provided
by vector $\vc{\tau}$ is indeed the capacity of the channel and
equals to:
\begin{equation}\label{eq: CapacityDC}
    C_{\text{DC}}=\frac{C_{01}C_{13}}{C_{01}+C_{13}}+\frac{C_{02}C_{23}}{C_{02}+C_{23}}.
\end{equation}
The result is valid for the Gaussian multiple antenna as well as
discrete memoryless channels, and therefore $\Delta=0$ ensures the
optimality of the MDF scheme for those channels too.
\subsection{MDF Gap Analysis}
To investigate how close the MDF scheme performs to the capacity of
the diamond channel when $\Delta\!\neq\!0$, the appropriate upper
bounds are required, which will be derived in sections \ref{sec:
MDF-BC} and \ref{sec: MDF-MAC}. Therefore, the detailed gap analysis
for the MDF scheme is deferred to Appendix \ref{appendix: MDF_Gap},
where it is shown that although a small gap is achievable for some
channel conditions, the gap can be large in general. In the
following sections, Broadcast and Multiple-Access Modes are added to
the MDF algorithm to achieve 0.71 bits of the capacity for
$\Delta\!>\!0$ and $\Delta\!<\!0$ cases, respectively.

\section{MDF-BC Scheme and Achieving within 0.71 Bits of the Capacity for $\Delta<0$}\label{sec: MDF-BC}
In the MDF scheme, since both branches cannot be fully utilized when
$\Delta\!<\!0$ simultaneously, there exists some unused capacity in
the second hop. To efficiently make use of the available resources,
Broadcast Mode is added to the MDF scheme. This mode provides the
relays with an additional reception time.
\subsection{Achievable Scheme}
The modified protocol uses Broadcast Mode together with Forward Modes I and II. Therefore, by setting $t_4=0$ in
Eq. (\ref{eq: Ach_Gen}) the maximum achievable rate of the scheme as a
function of the power allocation parameter $\eta$ used in superposition coding is:
\begin{equation*}
    R_{\text{BC}}(\eta)=\max_{\scriptstyle{\sum_{i=1}^{3}t_{i}= 1, t_{i}\geq 0}}\!\bigg\{\!\min\{t_{1}u(\eta)+t_{2}C_{01}, t_{3}C_{13}\}+ \min\{t_{1}v(\eta)+t_{3}C_{02},
    t_{2}C_{23}\!\}\bigg\}.
\end{equation*}
Recall that $u$, and $v$, defined respectively in Eqs. (\ref{eq:
u_alpha}) and (\ref{eq: v_alpha}), are the rates associated with
Relays 1 and 2 in Broadcast Mode. First, the optimal schedule is
obtained, assuming a fixed $\eta$, and later an appropriate value
for $\eta$ will be selected. The achievable rate can be written as
the following LP:
\begin{IEEEeqnarray}{ll}
\text{maximize} &R_{\textrm{BC}}\nonumber\\
\label{eq: BC-1}\text{subject to:}~&R_{\textrm{BC}} \leq t_{1}(u+v)+t_{2}C_{01}+t_{3}C_{02}\\
\label{eq: BC-2}&R_{\textrm{BC}} \leq t_{1}u+t_{2}(C_{01}+C_{23})+t_{3}.0\\
\label{eq: BC-3}&R_{\textrm{BC}} \leq t_{1}v+t_{2}.0+t_{3}(C_{02}+C_{13})\\
\label{eq: BC-4}&R_{\textrm{BC}} \leq t_{2}C_{23}+t_{3}C_{13}\\
\label{eq: BC-5}&\sum_{i=1}^{3} t_{i}= 1\\
\label{eq: BC-6}&t_{i}\geq0.
\end{IEEEeqnarray}
For a feasible LP, the solution is at one of the extreme points of
the constraint set. One of the extreme points can be obtained by
solving a set of linear equations containing Eq. (\ref{eq: BC-5})
and inequalities (\ref{eq: BC-1})-(\ref{eq: BC-3}) considered as
equalities. The solution becomes:
 \begin{IEEEeqnarray}{rl}
\label{eq: BC_t1}t_{1} &= \frac{-\Delta}{(C_{01}+C_{13})v+(C_{02}+C_{23})u-\Delta},\nonumber\\
t_{2} &=
\frac{C_{13}v+C_{02}u}{(C_{01}+C_{13})v+(C_{02}+C_{23})u-\Delta},\nonumber\\
t_{3} &=
\frac{C_{01}v+C_{23}u}{(C_{01}+C_{13})v+(C_{02}+C_{23})u-\Delta},\nonumber\\
\label{eq: RBC_Alpha}R_{\text{BC}}(\eta) &=
\frac{C_{13}(C_{01}+C_{23})v(\eta)+C_{23}(C_{02}+C_{13})u(\eta)}{(C_{01}+C_{13})v(\eta)+(C_{02}+C_{23})u(\eta)-\Delta}.
\end{IEEEeqnarray}

It is easy to verify
  that $\Delta\!=\!0$ makes $t_{1}\!=\!0$, and hence leads to the MDF algorithm. Note that in addition to inequalities
  (\ref{eq: BC-1})-(\ref{eq: BC-3}), the above extreme point also turns inequality (\ref{eq: BC-4})
into equality. Now, this extreme point is proven to be the solution
to the above LP. If one of the elements of vector $\textbf{t}$ is
increased,
 at least one of the conditions (\ref{eq: BC-1})-(\ref{eq: BC-3}) provides a smaller rate, compared to the rate obtained by the
 extreme point. For instance, if $t_{1}$ in Eq. (\ref{eq: BC_t1}) is increased, then, because of Eq. (\ref{eq: BC-5}),
  at least one of $t_{2}$ and $t_{3}$ should be decreased, which in turn reduces the rate associated with the inequality (\ref{eq: BC-4}). This confirms that the extreme point is the optimal solution to the LP with constraints (\ref{eq: BC-1})-(\ref{eq: BC-6}).

   In the following, instead of searching for $\eta^*$, which maximizes $R_{BC}(\eta)$, an appropriate value for $\eta$ is
   found that not only provides a small gap from the upper bounds, but also simplifies the gap analysis of section \ref{subsec: MDF-BC Gap}.
    The power allocation
parameter $\eta$ is selected to be either $\eta_{1}\triangleq\frac{1}{g_{01}+1}$, or $\eta_{2}\triangleq\frac{1}{g_{02}+1}$ for
 $C_{02}\geq C_{01}$ and $C_{01}\geq C_{02}$ conditions, respectively. As it will be shown in Appendix \ref{appendix: GDOF}, the chosen $\eta$ produces the same GDOF as the corresponding upper bound, which is a necessary condition in obtaining a small gap. The corresponding $u$ and $v$ for $\eta_1$ are:
\begin{IEEEeqnarray}{rl}
u(\eta_1)&=C_{01}-\zeta_1, \nonumber \\
\label{eq: uv_eta_1} v(\eta_1)&=C_{012}-C_{01},
\end{IEEEeqnarray}
and for $\eta_2$ are:
\begin{IEEEeqnarray}{rl}
u(\eta_2)&=C_{012}-C_{02}, \nonumber \\
\label{eq: uv_eta_2} v(\eta_2)&=C_{02}-\zeta_2.
\end{IEEEeqnarray}
In the above,
\begin{IEEEeqnarray}{rl}
  \label{eq: zeta_1}\zeta_{1} &\triangleq \mathcal{C}(\frac{g_{01}}{g_{01}+1})\leq \frac{1}{2}, \\
\label{eq: zeta}  \zeta_{2} &\triangleq \mathcal{C}(\frac{g_{02}}{g_{02}+1})\leq\frac{1}{2}.
\end{IEEEeqnarray}
The selected $\eta$ divides the source power between $u$ and $v$
(considered as the rates of two virtual users in the broadcast
channel consisting of ${\cal S}\!\Rightarrow\!{\cal R}_1$ and ${\cal
S}\!\Rightarrow\!{\cal R}_2$ links) in such a way that:
  \begin{enumerate}
    \item the sum data-rate (\emph{i.e.}, $u+v$) in the broadcast channel is close to the sum-capacity of the broadcast channel
    (\emph{i.e.}, $\max\{C_{01}, C_{02}\}$),
    \item the weaker user's rate is close to its capacity. For instance, if $C_{01}\leq C_{02}$, then $u\approx C_{01}$.
  \end{enumerate}
Substituting $u$ and $v$ from Eqs. (\ref{eq: uv_eta_1}) and
(\ref{eq: uv_eta_2}) into (\ref{eq: RBC_Alpha}) leads to the
following achievable rates $R_{\textrm{MDF-BC}}^{1}$ and
$R_{\textrm{MDF-BC}}^{2}$ corresponding to $\eta_{1}$ and
$\eta_{2}$:
 \begin{IEEEeqnarray}{rl}
 R_{\textrm{MDF-BC}}^{1}&=\frac{C_{13}(C_{01}+C_{23})C_{012}-C_{01}^{2}C_{13}+C_{01}C_{02}C_{23}-\zeta_{1} C_{23}(C_{02}+C_{13})}{(C_{01}+C_{13})(C_{012}-C_{01}+C_{23})-\zeta_{1}
    (C_{02}+C_{23})},\nonumber\\
     R_{\textrm{MDF-BC}}^{2}&=\frac{C_{23}(C_{02}+C_{13})C_{012}-C_{02}^{2}C_{23}+C_{01}C_{02}C_{13}-\zeta_{2} C_{13}(C_{01}+C_{23})}{(C_{02}+C_{23})(C_{012}-C_{02}+C_{13})-\zeta_{2}
    (C_{01}+C_{13})}.
    \end{IEEEeqnarray}
\subsection{Upper Bound}\label{subsec: MDF-BC_Cut-set Upper Bound}

   Following the discussion in section \ref{subsec: Cut-set Upper Bound}, we select one of the
   extreme points of the constraint set (\ref{eq: dual}) to obtain a fairly tight upper bound. Below, some insights on how to find an appropriate extreme point are given.

   First, Forward Modes I and II play an
   important role in data transfer from Source to Destination. These two modes let both Source
   and Destination be simultaneously active, which is important for efficient communication.
   This implies that \emph{generally} $t^{*}_{2}$ and $t^{*}_{3}$ are not zero in the original LP (\ref{eq: cut-set-org}).
   In addition, $\Delta\!<\!0$ roughly means that the second hop is
   better than the first hop. In this case, Broadcast Mode helps
   the relays to collect more data which will be sent to Destination using Forward Modes I and II later. Therefore, Multiple-Access Mode is less important when $\Delta\!<\!0$ and consequently $t_{4}$ can be set to zero. Using the complementary slackness
   theorem of linear programming (cf. \cite{book:LP}), having non-zero $t_{1}, t_{2}$, and $t_{3}$ in the original LP translates into
   having the first three inequalities in the dual program satisfied with equality.
   Now looking at the dual problem (\ref{eq: dual}) with the same structure as the original LP, in order to achieve a smaller
   objective function, we set $\tau_{2}$ or $\tau_{3}$ to zero. This is in contrast
   to the claim for having both of $t_{2}$ and $t_{3}$ non-zero in the original LP with the maximization objective.
   Therefore, the vector $\vc{\tau}$ with the
following properties is selected:
\begin{enumerate}
  \item Either $\tau_{2}$ or $\tau_{3}$ is zero.
  \item The first three inequalities are satisfied with equality.
\end{enumerate}
To have a valid $\vc{\tau}$, we need to make sure that all the
elements of vector $\vc{\tau}$ are non-negative and that $\vc{\tau}$
satisfies the last condition.

As mentioned earlier, either $\tau_{2}$ or $\tau_{3}$ can be set to zero in the dual program (\ref{eq: dual}). For instance, setting $\tau_{2}\!=\!0$ in the dual-program gives the following LP:
\begin{align}
\begin{array}{ll}\label{eq: Dual-LP-3var}
\text{minimize} &\widetilde{R} \\
\text{subject to:}~& \widetilde{R} \geq \tau_{1}C_{012}+\tau_{3}C_{02}+\tau_{4}.0 \\
&\widetilde{R} \geq \tau_{1}C_{01}+\tau_{3}.0+\tau_{4}C_{23} \\
&\widetilde{R} \geq \tau_{1}C_{02}+\tau_{3}(C_{02}+C_{13})+\tau_{4}C_{13} \\
&\widetilde{R} \geq \tau_{1}.0+\tau_{3}C_{13}+\tau_{4}C_{123} \\
&\sum_{i=1,i\neq2}^{4}\tau_{i}= 1,~\tau_{i}\geq 0.
\end{array}
\end{align}
Setting the first three inequalities to equalities gives:
\begin{IEEEeqnarray}{rl}
\tau_{1}^{*}&=\frac{C_{13}}{C_{012}-C_{02}+C_{13}},\nonumber\\
\tau_{3}^{*}&=\frac{C_{23}(C_{012}-C_{02})-C_{13}(C_{012}-C_{01})}{(C_{02}+C_{23})(C_{012}-C_{02}+C_{13})},\nonumber\\
\tau_{4}^{*}&=\frac{C_{13}(C_{012}-C_{01})+C_{02}(C_{012}-C_{02})}{(C_{02}+C_{23})(C_{012}-C_{02}+C_{13})},\nonumber\\
\label{eq:
up_instance}\widetilde{R}^{*}&=\frac{(C_{02}+C_{13})C_{23}}{C_{02}+C_{23}}+\frac{C_{13}(C_{01}C_{02}-C_{13}C_{23})}{(C_{02}+C_{23})(C_{012}-C_{02}+C_{13})}.
\end{IEEEeqnarray}
For obtaining a valid result, the following conditions have to be
ensured:
\begin{enumerate}
  \item $\tau_{3}^{*}\geq0$.
  \newline
   Since $C_{02}\!\leq\!C_{012}$, the denominator of $\tau_{3}^{*}$ is non-negative, therefore, the non-negativity of the nominator has to be guaranteed. This imposes the constraint $\Gamma\!\geq\!0$ on
  the values of channel parameters, where $\Gamma$ is defined as:
  \begin{equation}\label{eq: Gamma}
    \Gamma\triangleq C_{23}[C_{012}-C_{02}]-C_{13}[C_{012}-C_{01}].
  \end{equation}
  \item $\widetilde{R}^{*}\geq \tau_{3}^{*}C_{13}+\tau_{4}^{*}C_{123}$.
  \newline
  To satisfy the following condition:
  \begin{IEEEeqnarray*}{rl}
  \widetilde{R}^{*}=\tau_{1}^{*}C_{02}+\tau_{3}^{*}(C_{02}+C_{13})+\tau_{4}^{*}C_{13}&\geq \tau_{3}^{*}C_{13}+\tau_{4}^{*}C_{123},
  \end{IEEEeqnarray*}
  it is sufficient to show:
   \begin{IEEEeqnarray*}{rl}
  (\tau_{1}^{*}+\tau_{3}^{*})C_{02} &\geq \tau_{4}^{*}(C_{123}-C_{13}),
  \end{IEEEeqnarray*}
  which can be equivalently represented as:
  \begin{IEEEeqnarray*}{rl}
  \frac{C_{02}}{C_{123}-C_{13}+C_{02}} &\geq \tau_{4}^{*}.
  \end{IEEEeqnarray*}
   The following lemma proves the preceding inequality.
\end{enumerate}
\begin{lemma}\label{lemma1}
$\tau_{4}^{*}\leq \frac{C_{02}}{C_{123}-C_{13}+C_{02}}$ for
$C_{123}\leq C_{13}+C_{23}$.
\end{lemma}
\begin{IEEEproof}
See Appendix \ref{appendix1} .
\end{IEEEproof}
Lemma \ref{lemma1} requires $C_{123}\leq C_{13}+C_{23}$, which is
not true for $g_{13}g_{23}\leq4$. To be able to use Lemma
\ref{lemma1} for the case of  $g_{13}g_{23}\leq4$, we replace either
$C_{13}$ by $\hat{C}_{13}\triangleq C_{13}+\delta$ or $C_{23}$ by
$\hat{C}_{23}=C_{23}+\delta$ with $\delta$ defined as:
\begin{equation}\label{eq: delta}
    \delta\triangleq \max\{C_{123}-(C_{13}+C_{23}),0\}.
\end{equation}
This change provides the desired inequality (\emph{i.e.},
$C_{123}\leq \hat{C}_{13}+C_{23}$ or $C_{123}\leq
C_{13}+\hat{C}_{23}$) at the expense of increasing the upper bound.
However, we will show in Lemma \ref{lemma_up_dif} that this increase
is always less than $\delta$. We will prove that $\delta$ itself is
bounded in Lemma \ref{lemma_epsilon}.

Continuing the derivation of the upper bound from the LP (\ref{eq:
Dual-LP-3var}), if $C_{123}\!\geq\!C_{13}+C_{23}$, then $C_{23}$ is
replaced by $\hat{C}_{23}$. In this case, the dual program (\ref{eq:
Dual-LP-3var}) remains unchanged except for $C_{23}$. Hence, the set
of solutions (\ref{eq: up_instance}) can be used by replacing
$C_{23}$ with $\hat{C}_{23}$ and thus the upper bound becomes:
\begin{equation}\label{eq: R_increased}
    \hat{\widetilde{R}}^{*}=\frac{(C_{23}+\delta)(C_{02}+C_{13})}{C_{02}+C_{23}+\delta}+\frac{C_{13}\big(C_{01}C_{02}-C_{13}(C_{23}+\delta)\big)}{(C_{02}+C_{23}+\delta)(C_{012}-C_{02}+C_{13})}.
\end{equation}
Note that the inequality $\hat{\tau}_{3}^{*}\!\geq\!0$ holds because
$\hat{\Gamma}\!\geq\! 0$ simply follows from $\Gamma\!\geq\!
0$\footnote{The superscript $\hat{}$ is used to indicate parameters
associated with $\hat{C}_{23}$. For instance, $\hat{\Gamma}$ has the
same formula as $\Gamma$ in Eq. (\ref{eq: Gamma}), with $C_{23}$
replaced by $\hat{C}_{23}$.}. According to Lemma \ref{lemma1}, since
$C_{123}\!=\!C_{13}\!+\!\hat{C}_{23}$,
 the condition
$\hat{\widetilde{R}}^{*}\!\geq\!\hat{\tau}_{3}^{*}C_{13}\!+\!\hat{\tau}_{4}^{*}C_{123}$
is satisfied. Lemma
\ref{lemma_up_dif} shows that the enlarged upper bound $\hat{\widetilde{R}}^{*}$ (Eq. (\ref{eq: R_increased})) is at most $\delta$
bits greater than the upper bound of (\ref{eq: up_instance}).
\begin{lemma}\label{lemma_up_dif}
If $C_{123}\geq C_{13}+C_{23}$, then $\hat{\widetilde{R}}^{*}-\widetilde{R}^{*}\leq\delta$.
\end{lemma}
  \begin{IEEEproof}
  See Appendix \ref{appendix: lemma2}.
   \end{IEEEproof}
Therefore, the proposed upper bound for $\Delta\leq 0$ and $\Gamma>0$ is:
\begin{equation*}
    R_{\textrm{up}}^{2}=\frac{C_{23}(C_{02}+C_{13})}{C_{02}+C_{23}}+\frac{C_{13}\Delta}{(C_{012}-C_{02}+C_{13})(C_{02}+C_{23})}+\delta.
\end{equation*}

Similarly, when $\Delta\!\leq\!0$ and $\Gamma\!\leq\!0$, $\tau_{3}$ is set to zero and again the first three inequalities are assumed to be satisfied with equality
in the dual-program (\ref{eq: dual}). Following the same procedure,
the subsequent results are achieved:
\begin{IEEEeqnarray}{rl}
\tau_{1}^{*}&=\frac{C_{23}}{C_{012}-C_{01}+C_{23}},\nonumber\\
\tau_{2}^{*}&=\frac{C_{13}(C_{012}-C_{01})-C_{23}(C_{012}-C_{02})}{(C_{01}+C_{13})(C_{012}-C_{01}+C_{23})},\nonumber\\
\tau_{4}^{*}&=\frac{C_{23}(C_{012}-C_{02})+C_{01}(C_{012}-C_{01})}{(C_{01}+C_{13})(C_{012}-C_{01}+C_{23})},\nonumber\\
\widetilde{R}^{*}&=\frac{(C_{01}+C_{23})C_{13}}{C_{01}+C_{13}}+\frac{C_{23}(C_{01}C_{02}-C_{13}C_{23})}{(C_{01}+C_{13})(C_{012}-C_{01}+C_{23})},\nonumber\\
R_{\textrm{up}}^{1}&=\frac{C_{13}(C_{01}+C_{23})}{C_{01}+C_{13}}+\frac{C_{23}\Delta}{(C_{012}-C_{01}+C_{23})(C_{01}+C_{13})}+\delta.\label{eq: up_1}
\end{IEEEeqnarray} In this case, when
$C_{123}\geq C_{13}+C_{23}$, $\hat{C}_{13}$ is replaced by
$C_{13}+\delta$, it is easy to see that the preceding results can be
obtained by exchanging the roles of $C_{01}\leftrightarrow C_{02}$,
$C_{13}\leftrightarrow C_{23}$, and $\tau_{2}\leftrightarrow
\tau_{3}$ in the results derived for the case of $\Delta\leq 0$ and
$\Gamma>0$.

In order to be able to achieve a small gap from the upper bounds, $\delta$ should be bounded. Lemma
\ref{lemma_epsilon} proves that $\delta$ is smaller than 0.21
bits.
\begin{lemma} \label{lemma_epsilon}
   $\delta\leq\frac{1}{2}\log(\frac{4}{3})$.
   \end{lemma}
\begin{IEEEproof}
See Appendix \ref{appendix: lemma_epsilon}.
 \end{IEEEproof}
\subsection{Gap Analysis}\label{subsec: MDF-BC Gap}
The MDF-BC scheme is proposed for the
following regions:
\begin{enumerate}
  \item $\Delta<0$, $\Gamma\leq0$, $C_{02}\geq C_{01}$, and $C_{01}\geq 1$
  \item $\Delta<0$, $\Gamma\geq0$, $C_{01}\geq C_{02}$, and $C_{02}\geq 1$
\end{enumerate}
For $\Delta\!<\!0$, Appendix \ref{appendix: MDF_Gap} shows that the MDF
scheme provides a small gap from the upper bounds for the remaining regions. Here, the first case is considered.
The gap $\kappa_{\textrm{MDF-BC}}^{1}$ between the achievable rate $R_{\textrm{MDF-BC}}^{1}$ and the upper bound $R_{\textrm{up}}^{1}$ is:
\begin{equation*}\label{eq: gap_BC_2}
    \kappa_{\textrm{MDF-BC}}^{1}=\frac{-\zeta_{1}\bigg((C_{012}-C_{01}+C_{23})(C_{13}-C_{23})+C_{23}(C_{02}+C_{23})\bigg)\Delta}{(C_{01}+C_{13})(C_{012}-C_{01}+C_{23})\bigg((C_{01}+C_{13})(C_{012}-C_{01}+C_{23})-\zeta_{1}
    (C_{02}+C_{23})\bigg)}+\delta.
\end{equation*}
In the following lemma, the gap
$\kappa_{\textrm{MDF-BC}}^{1}$ is proved to be smaller than $\frac{1}{2}+\delta$
bits.
\begin{lemma}\label{lemma_C13C23}
$\kappa_{\textrm{MDF-BC}}^{1}\leq\frac{1}{2}+\delta$.\end{lemma}
\begin{IEEEproof}
See Appendix \ref{appendix: C13C23}.
\end{IEEEproof}

By exchanging the roles of $g_{01}\leftrightarrow g_{02}$ and
$g_{13}\leftrightarrow g_{23}$, the gap for the second case can be
easily derived and shown to be less than $\frac{1}{2}+\delta$ bits.

\section{MDF-MAC Scheme and Achieving within 0.71 Bits of the Capacity for $\Delta>0$}\label{sec: MDF-MAC}
Similar to section \ref{sec: MDF-BC}, a third mode is added to the
MDF scheme when $\Delta\!>\!0$ to effectively utilize the unused
capacity of the first hop.
\subsection{Achievable Scheme}
Here, Multiple-Access Mode is added to
the MDF scheme with independent messages sent from the relays to
Destination. This mode provides the relays with an increased
transmission time. The modified protocol uses three transmission
modes, \emph{i.e.}, Multiple-Access Mode and Forward Modes I and II. Therefore, by setting $t_{1}=0$ in Eq. (\ref{eq: Ach_Gen})
the maximum achievable rate of the scheme, $R_{\text{MAC}}$ is:
\begin{equation}\label{eq: Ach_MAC}
    R_{\textrm{MAC}}=\max_{\scriptstyle{\sum_{i=2}^{4}t_{i}= 1, t_{i}\geq 0}}\bigg\{\min\{t_{2}C_{01}, t_{3}C_{13}+R_{1}\}+ \min\{t_{3}C_{02},
    t_{2}C_{23}+R_{2}\}\bigg\},
\end{equation}
where $R_{1}$ and $R_{2}$ are the rates that Relays 1 and 2 provide
to Destination in Multiple-Access Mode, respectively. These rates
satisfy the multiple-access constraints in (\ref{eq: MAC-Cap-Reg}).
Lemma \ref{lemma: Ach_MAC} presents achievable rates, which will be
shown to be smaller than the capacity, by at most $.71$ bits, in
section \ref{subsec: MDF-MAC Gap}.
\begin{lemma}\label{lemma: Ach_MAC}
The achievable rates for $\Delta\!>\!0$ together with their corresponding scheduling are as follows:
\begin{IEEEeqnarray}{rl}
    R_{\textrm{MDF-MAC}}^{1}&=\frac{C_{01}(C_{02}+C_{13})}{C_{01}+C_{13}}-\frac{C_{02}\Delta}{(C_{01}+C_{13})(C_{\text{MAC}}-C_{13}+C_{02})}
    \quad \text{for} \quad \Delta>0, \ \Gamma'\leq 0,\nonumber\\
    \label{eq: MAC_Ach}
    R_{\textrm{MDF-MAC}}^{2}&=\frac{C_{02}(C_{01}+C_{23})}{C_{02}+C_{23}}-\frac{C_{01}\Delta}{(C_{02}+C_{23})(C_{\text{MAC}}-C_{23}+C_{01})}
    \quad \text{for} \quad \Delta>0, \ \Gamma'> 0,\label{eq:MAC_ACH}
\end{IEEEeqnarray}
\vspace{5pt}
  \begin{center}
  \begin{tabular}{|c|c|}
  \hline
  $\Gamma'\leq0$ & $\Gamma'>0$\\
  \hline
  $\begin{array}{rl}
  \vspace{5 pt}
t_{2}&=\frac{C_{13}}{C_{01}+C_{13}},\\
\vspace{5 pt}
t_{3}&=\frac{C_{01}(C_{\text{MAC}}-C_{13})+C_{13}C_{23}}{(C_{01}+C_{13})(C_{\text{MAC}}-C_{13}+C_{02})},\\
  \vspace{5 pt}
t_{4}&=\frac{\Delta}{(C_{01}+C_{13})(C_{\text{MAC}}-C_{13}+C_{02})},\\
\end{array}$ & $\begin{array}{rl}
  \vspace{5 pt}
t_{2}&=\frac{C_{02}(C_{\text{MAC}}-C_{23})+C_{13}C_{23}}{(C_{02}+C_{23})(C_{\text{MAC}}-C_{23}+C_{01})},\\
  \vspace{5 pt}
t_{3}&=\frac{C_{23}}{C_{02}+C_{23}},\\
  \vspace{5 pt}
t_{4}&=\frac{\Delta}{(C_{02}+C_{23})(C_{\text{MAC}}-C_{23}+C_{01})},
\end{array}$\\
 \hline
  \end{tabular}
  \end{center}
  where
\begin{equation}\label{eq: Gamma_prime}
   \Gamma'\triangleq C_{02}[C_{123}-C_{23}]-C_{01}[C_{123}-C_{13}].
\end{equation}
\end{lemma}
\begin{IEEEproof}
See Appendix \ref{appendix: Ach_MAC}.
\end{IEEEproof}

It is noted that if $\Delta=0$,
$t_{4}$ becomes zero and the scheme is converted to the MDF
scheme.
\subsection{Upper Bound}\label{subsec: MDF-MAC_Cut-set Upper Bound}
Following the same procedure as section \ref{subsec: MDF-BC_Cut-set
Upper Bound}, the upper bound for the case of  $\Delta\!\geq\!0$,
$\Gamma'\!\geq\!0$ is attained from (\ref{eq: up_instance}) by
exchanging the roles of $C_{01}\leftrightarrow C_{13}$,
$C_{02}\leftrightarrow C_{23}$, $\tau_{2}\leftrightarrow\tau_{3}$, and
$\tau_{1}\leftrightarrow\tau_{4}$. Similarly, when
$\Delta\!\geq\!0$ and $\Gamma'\!\leq\!0$, swapping the positions of
$C_{01}\leftrightarrow C_{23}$, $C_{02}\leftrightarrow C_{13}$, and
$\tau_{1}\leftrightarrow\tau_{4}$ in (\ref{eq: up_instance})
provides the upper bound. Therefore:
\begin{IEEEeqnarray}{rl}
R_{\textrm{up}}^{3}&=\frac{C_{01}(C_{02}+C_{13})}{C_{01}+C_{13}}+\frac{-C_{02}\Delta}{(C_{123}-C_{13}+C_{02})(C_{01}+C_{13})}+\delta \ \text{for $\Gamma'\leq 0$}, \nonumber\\
R_{\textrm{up}}^{4}&=\frac{C_{02}(C_{01}+C_{23})}{C_{02}+C_{23}}+\frac{-C_{01}\Delta}{(C_{123}-C_{23}+C_{01})(C_{02}+C_{23})}+\delta \ \text{for $\Gamma'> 0$}.\label{eq: up3_4}
\end{IEEEeqnarray}
\subsection{Gap Analysis}\label{subsec: MDF-MAC Gap}
By comparing the achievable rates (\ref{eq:MAC_ACH}) and the upper
bounds (\ref{eq: up3_4}), the gaps $\kappa_{\textrm{MAC}}^{1}$ and
$\kappa_{\textrm{MAC}}^{2}$ are respectively calculated for
$\Gamma'\!\leq\!0$ and $\Gamma'\!>\!0$ cases as:
\begin{IEEEeqnarray*}{rl}
\kappa_{\textrm{MAC}}^{1}\triangleq R_{\textrm{up}}^{3}-R_{\textrm{MDF-MAC}}^{1} &=\frac{ C_{02}(C_{123}\!-\!C_{\text{MAC}})\Delta}{(C_{01}\!+\!C_{13})(C_{\text{MAC}}\!-\!C_{13}\!+\!C_{02})(C_{123}\!-\!C_{13}\!+\!C_{02})}\!+\!\delta,\\
\kappa_{\textrm{MAC}}^{2}\triangleq R_{\textrm{up}}^{4}-R_{\textrm{MDF-MAC}}^{2}&=\frac{
C_{01}(C_{123}\!-\!C_{\text{MAC}})\Delta}{(C_{02}\!+\!C_{23})(C_{\text{MAC}}\!-\!C_{23}\!+\!C_{01})(C_{123}\!-\!C_{23}\!+\!C_{01})}\!+\!\delta.
\end{IEEEeqnarray*}

To show that the above gaps are small, Lemma \ref{lemma:
C123-CMAC} is employed.
\begin{lemma}\label{lemma: C123-CMAC}
$C_{123}\!-\!C_{\text{MAC}}\!\leq\!\frac{1}{2}$.
\end{lemma}
\begin{IEEEproof}
See Appendix \ref{appendix: lemma6}.
\end{IEEEproof}
Considering Lemma \ref{lemma: C123-CMAC}, it is straightforward to
show that the gap is at most $\frac{1}{2}\!+\!\delta$ bits.
Therefore, adding Multiple-Access Mode, with independent messages sent from the
relays to Destination, to the MDF scheme ensures the gap of less
than .71 bits from the upper bounds for $\Delta>0$.

\section{Conclusion}
In this work, we considered a dual-hop network with two parallel
relays in which each transmitting node has a constant power
constraint. We categorized the network into three classes based on
the fundamental parameter of the network $\Delta$, defined in this
paper. We derived explicit upper bounds for the different classes
using the cut-set bound. Based on the upper bounds, we proved that
the MDF scheme, which employs two transmission modes (Forward Modes
I and II), achieves the capacity of the channel when $\Delta\!=\!0$.
Furthermore, we analyzed
the gap between the achievable rate of the MDF scheme and the upper
bounds, showing that the gap can be large in some ranges of
parameters when $\Delta\!\neq\!0$. To guarantee the gap of at most
0.71 bits from the bounds, we added an extra broadcast or
multiple-access mode to the baseline MDF scheme for the cases of
$\Delta\!<\!0$ and $\Delta\!>\!0$, respectively. In addition, we
provided the asymptotic capacity analysis in the high SNR regime.
Finally, we argued that when the transmitting nodes operate under
average power constraints, the gap between the achievable scheme and
the cut-set upper bound is at most 3.6 bits.
\section*{Acknowledgment}
Helpful discussions with Mr. Oveis Gharan, especially on the proof of Appendix \ref{appendix: general average power} are acknowledged.

\appendices
\section{Generalized Degrees of Freedom Characterization}\label{appendix: GDOF}
It is interesting to consider the asymptotic capacity of the diamond channel in the high SNR regime.
A useful parameter in studying this capacity is the GDOF (cf. \cite{EtkinIT08, PhD:Avestimehr}) defined as:
\begin{equation*}
    \text{GDOF}(\vc{\alpha})\triangleq \lim_{P\rightarrow\infty}\frac{R}{\log P},
\end{equation*}
where $R$ is the data-rate, $P$ is a channel parameter (can be considered as SNR), and $\vc{\alpha}=\{\alpha_{01},\alpha_{02},\alpha_{13},\alpha_{23}\}$ with
\begin{IEEEeqnarray*}{rl}
    \alpha_{ij}&\triangleq \lim_{P\rightarrow\infty}\frac{\log(g_{ij})}{\log P} \quad \text{for} \ i \in\{0,1,2\}, \text{and} \ j\in\{1,2,3\}.
    \end{IEEEeqnarray*}

The vector $\vc{\alpha}$ shows how channel gains scale with $P$. Based on the above definition, the following approximations are valid:
\begin{IEEEeqnarray*}{rl}
C_{ij}&=\frac{1}{2}\log(1+g_{ij})\approx\frac{1}{2}\alpha_{ij}\log P,\\
C_{012}&=\frac{1}{2}\log(1+g_{01}+g_{02})\approx\frac{1}{2}\max\{\alpha_{01},\alpha_{02}\}\log P,\\
C_{123}&=\frac{1}{2}\log\big(1+(\sqrt{g_{13}}+\sqrt{g_{23}})^{2}\big)\approx\frac{1}{2}\max\{\alpha_{13},\alpha_{23}\}\log P,\\
C_{\text{MAC}}&=\frac{1}{2}\log(1+g_{13}+g_{23})\approx\frac{1}{2}\max\{\alpha_{13},\alpha_{23}\}\log P,\\
\Gamma&\approx \Big\{\alpha_{23}(\max\{\alpha_{01},\alpha_{02}\}-\alpha_{02})-\alpha_{13}(\max\{\alpha_{01},\alpha_{02}\}-\alpha_{01})\Big\}(\log P)^{2}+\sigma \log(P),\\
\Gamma'&\approx \Big\{\alpha_{02}(\max\{\alpha_{13},\alpha_{23}\}-\alpha_{23})-\alpha_{01}(\max\{\alpha_{13},\alpha_{23}\}-\alpha_{13})\Big\}(\log P)^{2}+\sigma' \log(P),
\end{IEEEeqnarray*}
where $\sigma$ and $\sigma'$ are positive constants. In the following analysis, it is assumed that $(\log P)^{2}$ terms are dominant,
 \emph{i.e.}, the coefficients of $(\log P)^{2}$ for $\Gamma$ and $\Gamma'$ are not zero. If this assumption is not valid, MDF scheme achieves the optimum GDOF of the channel.
According to the above approximations, it is easy to infer:
\begin{equation*}
\left\{
  \begin{array}{ll}
    \Gamma\leq0, & \hbox{if $\alpha_{01}\leq \alpha_{02}$;} \\
    \Gamma>0, & \hbox{if $\alpha_{01}> \alpha_{02}$;} \\
    \Gamma'\leq0, & \hbox{if $\alpha_{13}\leq \alpha_{23}$;} \\
    \Gamma'>0, & \hbox{if $\alpha_{13}> \alpha_{23}$.}
  \end{array}
\right.
\end{equation*}
Therefore, the GDOF associated with the upper bounds is:
\begin{IEEEeqnarray}{rl}
\text{GDOF}_{\textrm{up}}^{1}&=\frac{\alpha_{13}(\alpha_{01}+\alpha_{23})}{\alpha_{01}+\alpha_{13}}+\frac{\alpha_{23}(\alpha_{01}\alpha_{02}-\alpha_{13}\alpha_{23}) }{(\alpha_{01}+\alpha_{13})(\alpha_{02}-\alpha_{01}+\alpha_{23})},\nonumber\\
\text{GDOF}_{\textrm{up}}^{2}&=\frac{\alpha_{23}(\alpha_{02}+\alpha_{13})}{\alpha_{02}+\alpha_{23}}+\frac{\alpha_{13}(\alpha_{01}\alpha_{02}-\alpha_{13}\alpha_{23}) }{(\alpha_{02}+\alpha_{23})(\alpha_{01}-\alpha_{02}+\alpha_{13})},\nonumber\\
\text{GDOF}_{\textrm{up}}^{3}&=\frac{\alpha_{01}(\alpha_{02}+\alpha_{13})}{\alpha_{01}+\alpha_{13}}+\frac{-\alpha_{02}(\alpha_{01}\alpha_{02}-\alpha_{13}\alpha_{23}) }{(\alpha_{01}+\alpha_{13})(\alpha_{23}-\alpha_{13}+\alpha_{02})},\nonumber\\
\text{GDOF}_{\textrm{up}}^{4}&=\frac{\alpha_{02}(\alpha_{01}+\alpha_{23})}{\alpha_{02}+\alpha_{23}}+\frac{-\alpha_{01}(\alpha_{01}\alpha_{02}-\alpha_{13}\alpha_{23}) }{(\alpha_{02}+\alpha_{23})(\alpha_{13}-\alpha_{23}+\alpha_{01})}.
\end{IEEEeqnarray}

The GDOF for different achievablity schemes is as follows:

\textbf{MDF}:
\begin{align}
\begin{array}{ll}
\text{GDOF}_{\textrm{MDF}}^{1}&=\frac{\textstyle{\alpha_{01}(\alpha_{02}+\alpha_{13})}}{\textstyle{\alpha_{01}+\alpha_{13}}}, \vspace*{6pt}\\
\text{GDOF}_{\textrm{MDF}}^{2}&=\frac{\textstyle{\alpha_{02}(\alpha_{01}+\alpha_{23})}}{\textstyle{\alpha_{02}+\alpha_{23}}}, \vspace*{6pt}\\
\text{GDOF}_{\textrm{MDF}}^{3}&=\frac{\textstyle{\alpha_{13}(\alpha_{01}+\alpha_{23})}}{\textstyle{\alpha_{01}+\alpha_{13}}},\vspace*{6pt}\\
\text{GDOF}_{\textrm{MDF}}^{4}&=\frac{\textstyle{\alpha_{23}(\alpha_{02}+\alpha_{13})}}{\textstyle{\alpha_{02}+\alpha_{23}}}.
\end{array}
\end{align}

\textbf{MDF-BC}:
\begin{IEEEeqnarray}{rl}
\text{GDOF}_{\textrm{MDF-BC}}^{1}&=\frac{\alpha_{02}\alpha_{13}(\alpha_{01}+\alpha_{23})-\alpha_{01}^{2}\alpha_{13}+\alpha_{01}\alpha_{02}\alpha_{23}}{(\alpha_{01}+\alpha_{13})(\alpha_{02}-\alpha_{01}+\alpha_{23})},\nonumber\\
\text{GDOF}_{\textrm{MDF-BC}}^{2}&=\frac{\alpha_{01}\alpha_{23}(\alpha_{02}+\alpha_{13})-\alpha_{02}^{2}\alpha_{23}+\alpha_{01}\alpha_{02}\alpha_{13}}{(\alpha_{02}+\alpha_{23})(\alpha_{01}-\alpha_{02}+\alpha_{13})}.
\end{IEEEeqnarray}

\textbf{MDF-MAC}:
\begin{IEEEeqnarray}{rl}
\text{GDOF}_{\textrm{MDF-MAC}}^{1}&=\frac{\alpha_{01}(\alpha_{02}+\alpha_{13})}{\alpha_{01}+\alpha_{13}}-\frac{\alpha_{02}(\alpha_{01}\alpha_{02}-\alpha_{13}\alpha_{23})}{(\alpha_{01}+\alpha_{13})(\alpha_{23}-\alpha_{13}+\alpha_{02})},\nonumber\\
\text{GDOF}_{\textrm{MDF-MAC}}^{2}&=\frac{\alpha_{02}(\alpha_{01}+\alpha_{23})}{\alpha_{02}+\alpha_{23}}-\frac{\alpha_{01}(\alpha_{01}\alpha_{02}-\alpha_{13}\alpha_{23})}{(\alpha_{02}+\alpha_{23})(\alpha_{13}-\alpha_{23}+\alpha_{01})}.
\end{IEEEeqnarray}

By comparing the upper bounds on the GDOF and the achievable GDOFs,
it is easy to see that MDF-BC and MDF-MAC achieve the optimum GDOF
of the channel, while the MDF cannot achieve it for all channel
parameters.
\section{Diamond Channel with Average Power
Constraints}\label{appendix: general average power} In this appendix,
it is shown that if the transmitting nodes are subject to average power constraints, each of the cut-set bounds in Eq. (\ref{eq:
cut-set-org}) is increased at most by $\frac{2}{\ln 2}$ bits. This
analysis confirms that the achievable schemes proposed in this paper
with constant power constraints are still valid. In other
words, they provide a gap of at most $.71+\frac{2}{\ln 2}\leq 3.6$
bits from the cut-set bounds.

Let $P_{{\cal S}}^{(i)*}$, $P_{{\cal R}_1}^{(i)*}$, and $P_{{\cal
R}_2}^{(i)*}$, for $i\in\{1,\cdots,4\}$ be the optimum power
allocated to Source, Relay 1, and Relay 2 in transmission mode $i$
with the corresponding time interval $t^{*}_{i}$ leading to the
cut-set bound $R_0$. The following constraints are in
effect\footnote{For the purpose of clarity, here the average powers
are not set to unity.}:
\begin{IEEEeqnarray}{rl}\label{eq: average_power}
  \sum_{i=1}^{4} t^{*}_{i}P_{{\cal S}}^{(i)*} &\leq P_{{\cal S}}, \nonumber \\
  \sum_{i=1}^{4} t^{*}_{i}P_{{\cal R}_1}^{(i)*} &\leq P_{{\cal R}_1}, \\
  \sum_{i=1}^{4} t^{*}_{i}P_{{\cal R}_2}^{(i)*} &\leq P_{{\cal R}_2}. \nonumber
  \end{IEEEeqnarray}
Therefore, the cut-set upper bound $R_0$ satisfies the following constraints:
\begin{equation}\label{eq: cut-set-avg}
\begin{array}{ll}
R_0 &\leq t^{*}_{1}\mathcal{C}\big((g_{01}+g_{02})P_{{\cal S}}^{(1)*}\big)+t^{*}_{2}\mathcal{C}(g_{01}P_{{\cal S}}^{(2)*})+t^{*}_{3}\mathcal{C}(g_{02}P_{{\cal S}}^{(3)*}),\\
R_0 &\leq t^{*}_{1}\mathcal{C}(g_{01}P_{{\cal S}}^{(1)*})+t^{*}_{2}\Big(\mathcal{C}(g_{01}P_{{\cal S}}^{(2)*})+\mathcal{C}(g_{23}P_{{\cal R}_2}^{(2)*})\Big)+t^{*}_{4}\mathcal{C}(g_{23}P_{{\cal R}_2}^{(4)*}) ,\\
R_0 &\leq t^{*}_{1}\mathcal{C}(g_{02}P_{{\cal S}}^{(1)*})+t^{*}_{3}\Big(\mathcal{C}(g_{02}P_{{\cal S}}^{(3)*})+\mathcal{C}(g_{13}P_{{\cal R}_1}^{(3)*})\Big)+t^{*}_{4}\mathcal{C}(g_{13}P_{{\cal R}_1}^{(4)*}) ,\\
R_0 &\leq t^{*}_{2}\mathcal{C}(g_{23}P_{{\cal R}_2}^{(2)*})+t^{*}_{3}\mathcal{C}(g_{13}P_{{\cal R}_1}^{(3)*})+t^{*}_{4}\mathcal{C}\Big(\big(\sqrt{g_{13}P_{{\cal R}_1}^{(4)*}}+\sqrt{g_{23}P_{{\cal R}_2}^{(4)*}}\big)^{2}\Big).
\end{array}
\end{equation}

Suppose that the vector $\textbf{t}'$ is the solution to the LP
(\ref{eq: cut-set-org}) leading to the rate $R_{1}$. If the vector
$\textbf{t}^{*}$ is used instead of $\textbf{t}'$ in the LP
(\ref{eq: cut-set-org}), the resulting rate that satisfies the
conditions of the LP, called $R_{2}$, becomes smaller than $R_{1}$.
It is clear that the increase in the cut-set bound due to the
\emph{average} instead of the \emph{constant} power constraints
(compare Eq. (\ref{eq: constant_power}) to Eq. (\ref{eq:
average_power})), \emph{i.e.}, $R_0-R_{1}$ is smaller than
$R_0-R_{2}$. Here, it is proved that $R_0-R_{2}\leq
\frac{2}{\ln{2}}$.

Consider each component term in the form of
$t^{*}_{i}\mathcal{C}(.)$ present in the inequality set (\ref{eq:
cut-set-avg}). For instance, consider $R_{c,0}\triangleq
t^{*}_{1}\mathcal{C}(g_{02}P_{{\cal S}}^{(1)*})$. The corresponding
term in constructing $R_{2}$ is $R_{c,2}\triangleq
t^{*}_{1}\mathcal{C}(g_{02}P_{{\cal S}})$. Because of the power
constraints (\ref{eq: average_power}), $R_{c,0}\leq
t^{*}_{1}\mathcal{C}(g_{02}\frac{P_{{\cal S}}}{t^{*}_{1}})$.
Therefore, it is easy to show:
\begin{IEEEeqnarray*}{rl}
R_{c,0}-R_{c,2}&\leq t^{*}_{1}\mathcal{C}\big(\frac{g_{02}P_{{\cal S}}(1-t^{*}_{1})}{(1+g_{02}P_{{\cal S}})t^{*}_{1}}\big)\\
{}&\stackrel{(a)}{\leq} \frac{g_{02}P_{{\cal S}}(1-t^{*}_{1})}{2(1+g_{02}P_{{\cal S}})\ln 2}\\
{}&\leq \frac{1}{2 \ln 2},
\end{IEEEeqnarray*}
where $(a)$ is due to the fact that $\mathcal{C}(x)\leq
\frac{x}{2\ln 2}$ for any $x\geq 0$. Similar analysis applies to
each component term. It is observed that the first and fourth
cut-set bounds in inequality set (\ref{eq: cut-set-avg}) have three
component terms and the second and third cut-set bounds have four
component terms. Therefore, $R_0-R_{2}\leq \frac{2}{\ln{2}}$.
\section{MDF Gap Analysis}\label{appendix: MDF_Gap}
We investigate how close the MDF scheme performs to the upper bounds
when $\Delta\neq0$. First, the gap between the MDF scheme and the
upper bound is calculated for regions specified in Table I. Then,
two special cases are considered.

\textbf{\emph{General Case}.}
We calculate the difference, named
$\kappa$, between the upper bounds and the rate offered by the MDF scheme from Eq. (\ref{eq: MDF_rate})
for the cases shown in Table I (see Appendix \ref{appendix Table}):

\begin{align*}\label{eq: GAP_4}
    \kappa_{1}&=\frac{-(C_{012}-C_{01})\Delta}{(C_{01}+C_{13})(C_{012}-C_{01}+C_{23})}+\delta,\nonumber\\
\kappa_{2}&=\frac{-(C_{012}-C_{02})\Delta}{(C_{02}+C_{23})(C_{012}-C_{02}+C_{13})}+\delta,\nonumber\\
\kappa_{3}&=\frac{(C_{123}-C_{13})\Delta}{(C_{01}+C_{13})(C_{123}-C_{13}+C_{02})}+\delta,\nonumber\\
\kappa_{4}&=\frac{(C_{123}-C_{23})\Delta}{(C_{02}+C_{23})(C_{123}-C_{23}+C_{01})}+\delta,\nonumber\\
\kappa_{5}&=\frac{-\Delta}{C_{01}+C_{13}}
\Big(\frac{C_{01}+C_{23}}{C_{02}+C_{23}}-\frac{C_{23}}{C_{012}-C_{01}+C_{23}}\Big)+\delta,\nonumber\\
\kappa_{6}&=\frac{-\Delta}{C_{02}+C_{23}}
\Big(\frac{C_{02}+C_{13}}{C_{01}+C_{13}}-\frac{C_{13}}{C_{012}-C_{02}+C_{13}}\Big)+\delta,\nonumber\\
\kappa_{7}&=\frac{\Delta}{C_{01}+C_{13}}
\Big(\frac{C_{02}+C_{13}}{C_{02}+C_{23}}-\frac{C_{02}}{C_{123}-C_{13}+C_{02}}\Big),\nonumber\\
\kappa_{8}&=\frac{\Delta}{C_{02}+C_{23}}
\Big(\frac{C_{01}+C_{23}}{C_{01}+C_{13}}-\frac{C_{01}}{C_{123}-C_{23}+C_{01}}\Big).\nonumber
\end{align*}
Note that for the regions associated with $\kappa_{7}$ and
$\kappa_{8}$ specified in Table I, $C_{123}\leq C_{13}+C_{23}$ and
hence, $\delta=0$.

To prove that $\kappa_{i}$ for $i\in\{1,\cdots,4\}$ are small, the
following lemma is needed:
\begin{lemma}\label{lemma: DiffC012}
\begin{IEEEeqnarray*}{ll}
C_{012}-\max\{C_{01},C_{02}\}&\leq \frac{1}{2},\\
C_{123}-\max\{C_{13},C_{23}\}&\leq 1.
\end{IEEEeqnarray*}
\end{lemma}
\begin{IEEEproof}
See Appendix \ref{appendix2}.
 \end{IEEEproof}

 For instance, following $\kappa_{1}\leq \frac{1}{2}+\delta$ is proved:
\begin{IEEEeqnarray*}{ll}
    \kappa_{1}&=\frac{(C_{13}C_{23}-C_{01}C_{02})(C_{012}-C_{01})}{(C_{01}+C_{13})(C_{012}-C_{01}+C_{23})}+\delta\nonumber\\
    {}&\stackrel{(a)}{\leq} \frac{C_{13}C_{23}(C_{012}-C_{01})}{(C_{01}+C_{13})(C_{012}-C_{01}+C_{23})}+\delta\nonumber\\
    {}&\stackrel{(b)}{\leq} \frac{1}{2}\frac{C_{13}C_{23}}{(C_{01}+C_{13})(C_{012}-C_{01}+C_{23})}+\delta\nonumber\\
    {}&=\frac{1}{2}\frac{C_{13}}{C_{01}+C_{13}}\times \frac{C_{23}}{C_{012}-C_{01}+C_{23}}+\delta\nonumber\\
    {}&\leq \frac{1}{2}+\delta,
 \end{IEEEeqnarray*}
where $(a)$ comes from the fact that $\Delta\!>\!0$ for this case.
According to the corresponding region shown in Table I,
$C_{02}\!\leq\!C_{01}$ and therefore $(b)$ is true based on Lemma
\ref{lemma: DiffC012}.

Lemmas \ref{lemma: kappa5} and \ref{lemma: kappa7} prove that
$\kappa_{5}\!\leq\!\frac{1}{2}\!+\!\delta$ and
$\kappa_{7}\!\leq\!1$, respectively. The proof techniques can be
easily adopted to correspondingly show that
$\kappa_{6}\!\leq\!\frac{1}{2}\!+\!\delta$, and
$\kappa_{8}\!\leq\!1$.
\begin{lemma}\label{lemma: kappa5}
$\kappa_{5}\leq\frac{1}{2}+\delta$.
\end{lemma}
\begin{IEEEproof}
See Appendix \ref{appendix: kappa5}.
\end{IEEEproof}
\begin{lemma}\label{lemma: kappa7}
$\kappa_{7}\leq 1.$
\end{lemma}
\begin{IEEEproof}
See Appendix \ref{appendix: kappa7}.
\end{IEEEproof}
Two special cases are also considered:

\textbf{\emph{Symmetric Case}.} When $C_{01}\!=\!C_{02}$ and
$C_{13}\!=\!C_{23}$, $\Gamma\!=\!\Gamma'\!=\!0$ and it can be seen
from Table I that the MDF scheme offers a data-rate that is, at
most, $1+\delta$ bits less than the corresponding upper bound.

\textbf{\emph{Partially Symmetric Case}.} When either
$C_{01}\!=\!C_{02}$ with $\Delta\!<\!0$, or $C_{13}\!=\!C_{23}$ with
$\Delta\!>\!0$ occurs, it was seen in section \ref{subsection:
MDF-Ach} that fully utilizing branch 1 or branch 2 gives the same
achievable rate. Table I shows that in such cases, the gap is less
than $1+\delta$ bits.

\textbf{\emph{Discussion}.} Multiplexing Gain (MG) of a scheme is
defined in \cite{IT08_1:Sreeram, ISIT05:Host_Madsen} as:
\begin{equation*}
    \text{MG}\triangleq \lim_{\text{SNR}\rightarrow \infty} \frac{R}{0.5\log(\text{SNR})},
\end{equation*}
where $R$ is the achievable rate of the scheme. Using Eq. (\ref{eq:
MDF_rate}), it can be shown that the MDF scheme achieves the
multiplexing gain of 1. Avestimehr, \emph{et.al} proposed a
broadcast mutiple-access scheme for the full-duplex diamond channel
and proved that the scheme is within one bit from the cut-set bound
\cite{ISIT08:Avestimehr}. In the half-duplex case, the multiplexing
gain of 1 is lost if this approach is followed, leading to an
infinite gap between the achievable rate and the upper bound.

It is easy to show that, for the remaining cases (shown in Table I),
the gap can be large. For instance, suppose $C_{02}\!=\!x$,
$C_{13}\!=\!C_{23}\!=\!\alpha x$ and $C_{01}\!=\!\beta x$, with
$\alpha\!>\!\beta\!>\!1$. In this case $\Delta\!<\!0$, and
$\Gamma\!>\!0$ and therefore, the gap $\kappa$ is:
\begin{IEEEeqnarray*}{ll}
\kappa&=\frac{-\Delta}{C_{02}+C_{23}}
\Big(\frac{C_{02}+C_{13}}{C_{01}+C_{13}}-\frac{C_{13}}{C_{012}-C_{02}+C_{13}}\Big)+\delta\nonumber\\
{}&=\frac{-\Delta}{C_{02}+C_{23}}
\Big(\frac{C_{02}(C_{012}-C_{02})+C_{13}(C_{012}-C_{01})}{(C_{01}+C_{13})(C_{012}-C_{02}+C_{13})}\Big)+\delta\nonumber\\
{}&\stackrel{(a)}{\geq} \frac{-\Delta}{C_{02}+C_{23}}
\Big(\frac{C_{02}(C_{012}-C_{02})}{(C_{01}+C_{13})(C_{012}-C_{02}+C_{13})}\Big)+\delta\nonumber\\
{}&\stackrel{(b)}{\geq} \frac{-\Delta}{C_{02}+C_{23}}
\Big(\frac{C_{02}(C_{01}-C_{02})}{(C_{01}+C_{13})(C_{012}-C_{02}+C_{13})}\Big)+\delta\nonumber\\
{}&\stackrel{(c)}{\geq} \frac{-\Delta}{C_{02}+C_{23}}
\Big(\frac{C_{02}(C_{01}-C_{02})}{(C_{01}+C_{13})^{2}}\Big)+\delta\nonumber\\
{}&\stackrel{(d)}{=}\frac{(\alpha^{2}-\beta)(\beta-1)}{(\alpha+\beta)^{2}(\alpha+1)}\
x+\delta,
\end{IEEEeqnarray*}
where in $(a)$ the nominator is decreased by
$C_{13}(C_{012}-C_{01})$. To obtain $(b)$, $C_{012}$ in the
nominator is replaced by the smaller quantity $C_{01}$. For $(c)$,
$C_{012}$ is substituted by the larger term $C_{01}+C_{02}$ in the
denominator. In $(d)$, the assumed values of the capacities in terms
of $x$ are substituted. It is clear that the gap increases as
$x$ becomes large. GDOF analysis of Appendix \ref{appendix: GDOF} also confirms that the MDF scheme can have a large gap from the upper bound.
\section{Proofs} \label{appendix proofs}
In this appendix, the proofs of the lemmas used in this paper are provided.
\subsection{Proof of Lemma \ref{lemma1}}\label{appendix1} We start
with the fact that $C_{01}+C_{02}\!\geq\!C_{012}$. Rearranging the
terms, and multiplying both sides of the inequality by $C_{13}$
give:
\begin{equation*}
    C_{13}C_{02} \geq C_{13}(C_{012}-C_{01}).
\end{equation*}
By adding $C_{02}(C_{012}-C_{02})$ to both
sides and then dividing both sides by $C_{012}\!-C_{02}\!+\!C_{13}$, we obtain:
\begin{equation*}
    C_{02}\geq
\frac{C_{13}(C_{012}-C_{01})+C_{02}(C_{012}-C_{02})}{C_{012}-C_{02}+C_{13}}.
\end{equation*}
Assuming $C_{123}\leq C_{13}+C_{23}$, we
divide the Right Hand Side (RHS)
by $C_{02}+C_{23}$ and the Left Hand Side (LHS) by the smaller quantity
$C_{123}-C_{13}+C_{02}$ to achieve:
\begin{IEEEeqnarray*}{rl}
\frac{C_{02}}{C_{123}-C_{13}+C_{02}}&\geq
\frac{C_{13}(C_{012}-C_{01})+C_{02}(C_{012}-C_{02})}{(C_{012}-C_{02}+C_{13})(C_{02}+C_{23})}=\tau_{4}^{*}.
\end{IEEEeqnarray*}
This completes the proof.
\subsection{Proof of Lemma \ref{lemma_up_dif}}\label{appendix: lemma2}
  \begin{IEEEeqnarray*}{rl}
\hat{\widetilde{R}}^{*}-\widetilde{R}^{*}&=\frac{\delta
C_{02}\Big((C_{02}+C_{13})(C_{012}-C_{02}+C_{13})-C_{13}(C_{01}+C_{13})\Big)}{(C_{02}+C_{23})(C_{02}+C_{23}+\delta)(C_{012}-C_{02}+C_{13})}\nonumber\\
{}&\stackrel{(a)}{\leq} \frac{\delta
C_{02}^{2}}{(C_{02}+C_{23})^{2}}\nonumber\\
{}&\leq \delta,
\end{IEEEeqnarray*}
where in $(a)$, the nominator is increased by replacing
$C_{012}-C_{02}$ with $C_{01}$, using the fact that
$C_{012}-C_{02}\leq C_{01}$ (see Eq. (\ref{eq: C012})). In addition, the denominator is decreased by removing
$\delta$.
\subsection{Proof of Lemma \ref{lemma_epsilon}}\label{appendix: lemma_epsilon}
\begin{IEEEeqnarray}{rl}
\delta&=C_{123}-(C_{13}+C_{23})\nonumber\\
{}&=\frac{1}{2}\log\left(\frac{1+g_{13}+g_{23}+2\sqrt{g_{13}g_{23}}}{1+g_{13}+g_{23}+g_{13}g_{23}}\right)\nonumber\\
{}&\stackrel{(a)}{\leq}\frac{1}{2}\log\left(1+\frac{2\sqrt{g_{13}g_{23}}-g_{13}g_{23}}{1+2\sqrt{g_{13}g_{23}}+g_{13}g_{23}}\right)\nonumber\\
{}&\stackrel{(b)}{\leq} \frac{1}{2}\log(\frac{4}{3}),\nonumber
\end{IEEEeqnarray}
where in $(a)$ the denominator is decreased by replacing
$g_{13}+g_{23}$ with the smaller term  $2\sqrt{g_{13}g_{23}}$. Defining $x\triangleq \sqrt{g_{13}g_{23}}$, it is easy to show that the maximum
of $\log(1+\frac{2x-x^{2}}{1+2x+x^{2}})$, for $0\!\leq\! x\!\leq\! 2$, is $x^{*}=\frac{1}{2}$, \emph{i.e.},
$g_{13}^{*}g^{*}_{23}=\frac{1}{4}$, which proves $(b)$.
\subsection{Proof of Lemma \ref{lemma_C13C23}}\label{appendix: C13C23}
It is known that $C_{01}, C_{02} \leq C_{012}$, which proves $0 \leq
C_{23}(C_{012}-C_{02})$ and $0 \leq C_{01}(C_{012}-C_{01})$. Since
both terms are positive, the sum of them is also positive,
\emph{i.e.}, $0 \leq C_{23}(C_{012}-C_{02}) +
C_{01}(C_{012}-C_{01})$. By adding and subtracting
$(C_{012}-C_{01}+C_{23})C_{13}+C_{01}C_{13}$, the inequality can be
rearranged to:
\begin{equation*}
    0 \leq (C_{012}-C_{01}+C_{23})(C_{01}+C_{13}) + (C_{012}-C_{01})(C_{23}-C_{13})-C_{23}(C_{02}+C_{13}).
\end{equation*}
As mentioned earlier, Broadcast Mode is used for $\Delta\leq 0$,
\emph{i.e.}, $C_{01}C_{02}\leq C_{13}C_{23}$. Therefore, both sides are multiplied by the positive term $-\Delta$ to acquire:
\begin{equation*}
    0 \leq (C_{13}C_{23}-C_{01}C_{02}) \big( (C_{012}-C_{01}+C_{23})(C_{01}+C_{13}) + (C_{012}-C_{01})(C_{23}-C_{13})-C_{23}(C_{02}+C_{13})\big).
\end{equation*}
Now, the positive term $(C_{012}-C_{01}+C_{23})(C_{012}-C_{01})(C_{01}+C_{13})^{2}$ can be added to the RHS of the inequality to achieve:
\begin{IEEEeqnarray*}{rl}
    0 \leq &(C_{13}C_{23}-C_{01}C_{02}) \big( (C_{012}-C_{01}+C_{23})(C_{01}+C_{13})+ (C_{012}-C_{01})(C_{23}-C_{13})-C_{23}(C_{02}+C_{13})\big)\\
    {}&+(C_{012}-C_{01}+C_{23})(C_{012}-C_{01})(C_{01}+C_{13})^{2}.
\end{IEEEeqnarray*}
The above inequality can be equivalently stated as:
\begin{IEEEeqnarray*}{rl}
    (C_{13}C_{23}-C_{01}C_{02}) \Big( (C_{012}-C_{01}+C_{23})(C_{13}-C_{23})+ C_{23}(C_{02}+C_{23})\Big)&+\\ {} C_{01}(C_{02}+C_{23})(C_{01}+C_{13})(C_{012}-C_{01}+C_{23}) &\leq (C_{012}-C_{01}+C_{23})^{2}(C_{01}+C_{13})^{2}.
\end{IEEEeqnarray*}
Since $1 \leq C_{01}$, the LHS becomes smaller if $C_{01}(C_{02}+C_{23})$ is replaced by $(C_{02}+C_{23})$, leading to:
\begin{IEEEeqnarray*}{rl}
    (C_{13}C_{23}-C_{01}C_{02}) \Big( (C_{012}-C_{01}+C_{23})(C_{13}-C_{23})+ C_{23}(C_{02}+C_{23})\Big)&+\\ {} (C_{02}+C_{23})(C_{01}+C_{13})(C_{012}-C_{01}+C_{23}) &\leq (C_{012}-C_{01}+C_{23})^{2}(C_{01}+C_{13})^{2}.
\end{IEEEeqnarray*}
Now as $\zeta_{1}\leq \frac{1}{2}$ (see Eq. (\ref{eq: zeta})), the following inequality is also true:
\begin{IEEEeqnarray*}{rl}
    &\zeta_{1} \Bigg\{2\times (C_{13}C_{23}-C_{01}C_{02}) \Big( (C_{012}-C_{01}+C_{23})(C_{13}-C_{23})+ C_{23}(C_{02}+C_{23})\Big)+\\
    &{} (C_{02}+C_{23})(C_{01}+C_{13})(C_{012}-C_{01}+C_{23})\Bigg\} \leq (C_{012}-C_{01}+C_{23})^{2}(C_{01}+C_{13})^{2}.
\end{IEEEeqnarray*}
By rearranging the preceding inequality
\begin{IEEEeqnarray*}{rl}
    \frac{\zeta_{1} (C_{13}C_{23}-C_{01}C_{02}) \Big( (C_{012}-C_{01}+C_{23})(C_{13}-C_{23})+ C_{23}(C_{02}+C_{23})\Big)}{(C_{01}+C_{13})(C_{012}-C_{01}+C_{23})\Big((C_{01}+C_{13})(C_{012}-C_{01}+C_{23})-\zeta_{1}     (C_{02}+C_{23})\Big)} &\leq \frac{1}{2},
\end{IEEEeqnarray*}
which completes the proof.
\subsection{Proof of Lemma \ref{lemma: Ach_MAC}} \label{appendix: Ach_MAC} The optimization
(\ref{eq: Ach_MAC}) is an LP and together with the multiple-access
constraints (\ref{eq: MAC-Cap-Reg}) can be written as follows:
\begin{IEEEeqnarray*}{ll}
\text{maximize} &R_{\textrm{MAC}} \nonumber\\
\text{subject to:}~& R_{\textrm{MAC}} \leq t_{2}C_{01}+t_{3}C_{02}\\
&R_{\textrm{MAC}}-R_{1} \leq t_{3}(C_{02}+C_{13})\\
&R_{\textrm{MAC}}-R_{2} \leq t_{2}(C_{01}+C_{23})\\
&R_{\textrm{MAC}}-(R_{1}+R_{2}) \leq t_{2}C_{23}+t_{3}C_{13}\\
&R_{1}\leq t_{4}C_{13}\\
&R_{2}\leq t_{4}C_{23}\\
&R_{1}+R_{2}\leq t_{4}C_{\text{MAC}}\\
&\sum_{i=2}^{4} t_{i}= 1,~t_{i}\geq 0.
\end{IEEEeqnarray*}

Using Fourier-Motzkin elimination \cite{book:LP}, the LP can be equivalently stated as:
\begin{IEEEeqnarray}{ll}
\text{maximize} &R_{\textrm{MAC}} \nonumber\\
\label{eq: MAC-1}\text{subject to:}~& R_{\textrm{MAC}} \leq t_{2}C_{01}+t_{3}C_{02}\\
\label{eq: MAC-0}&R_{\textrm{MAC}} \leq t_{3}(C_{02}+C_{13})+t_{4}C_{13}\\
&R_{\textrm{MAC}}\leq t_{2}(C_{01}+C_{23})+t_{4}C_{23}\\
\label{eq: MAC-2}&R_{\textrm{MAC}}\leq t_{2}C_{23}+t_{3}C_{13}+t_{4}C_{\text{MAC}}\\
\label{eq: MAC-3}&R_{\textrm{MAC}}\leq t_{2}C_{23}+t_{3}C_{13}+t_{4}(C_{13}+C_{23})\\
\label{eq: MAC-4}&2R_{\textrm{MAC}}\leq t_{2}(C_{01}+C_{23})+t_{3}(C_{02}+C_{13})+t_{4}C_{\text{MAC}}\\
\label{eq: MAC-5}&2R_{\textrm{MAC}}\leq t_{2}C_{23}+t_{3}(C_{02}+2C_{13})+t_{4}(C_{13}+C_{\text{MAC}})\\
&\sum_{i=2}^{4} t_{i}= 1,~t_{i}\geq 0.
\end{IEEEeqnarray}

Now, it is shown that inequalities (\ref{eq: MAC-3})-(\ref{eq:
MAC-5}) are redundant. First, since $C_{\text{MAC}}\leq
(C_{13}+C_{23})$, the RHS of inequality (\ref{eq: MAC-3}) is greater
than the RHS of inequality (\ref{eq: MAC-2}). Therefore, inequality
(\ref{eq: MAC-3}) is redundant. Second, inequalities (\ref{eq:
MAC-4}) and (\ref{eq: MAC-5}) are simply obtained by adding
inequalities (\ref{eq: MAC-1}, \ref{eq: MAC-2}) and (\ref{eq:
MAC-0}, \ref{eq: MAC-2}), respectively. Therefore, the following LP
gives the maximum achievable rate of this scheme:
\begin{IEEEeqnarray}{ll}
\text{maximize} &R_{\textrm{MAC}} \nonumber\\
\label{eq: mac-1}\text{subject to:}~& R_{\textrm{MAC}} \leq t_{2}C_{01}+t_{3}C_{02}\\
\label{eq: mac-2}&R_{\textrm{MAC}} \leq t_{3}(C_{02}+C_{13})+t_{4}C_{13}\\
\label{eq: mac-3}&R_{\textrm{MAC}}\leq t_{2}(C_{01}+C_{23})+t_{4}C_{23}\\
\label{eq: mac-4}&R_{\textrm{MAC}}\leq t_{2}C_{23}+t_{3}C_{13}+t_{4}C_{\text{MAC}}\\
&\sum_{i=2}^{4} t_{i}= 1,~t_{i}\geq 0.
\end{IEEEeqnarray}

Instead of solving the above LP, a feasible solution that satisfies
all the constraints is found. This solution is not necessarily
optimum, however it provides us with an achievable rate. For
$\Gamma'\!\leq\!0$ inequalities (\ref{eq: mac-1}), (\ref{eq:
mac-2}), and (\ref{eq: mac-4}) are set to equalities, leading to:
\begin{IEEEeqnarray}{rl}\label{eqnarr: RMAC}
t_{2}&=\frac{C_{13}}{C_{01}+C_{13}},\nonumber\\
t_{3}&=\frac{C_{01}(C_{\text{MAC}}-C_{13})+C_{13}C_{23}}{(C_{01}+C_{13})(C_{\text{MAC}}-C_{13}+C_{02})},\nonumber\\
t_{4}&=\frac{\Delta}{(C_{01}+C_{13})(C_{\text{MAC}}-C_{13}+C_{02})},\nonumber\\
R_{\textrm{MDF-MAC}}^{1}&=\frac{C_{01}(C_{02}+C_{13})}{C_{01}+C_{13}}-\frac{ C_{02}\Delta}{(C_{01}+C_{13})(C_{\text{MAC}}-C_{13}+C_{02})}.
\end{IEEEeqnarray}

To ensure that the above results are valid, the inequality (\ref{eq: mac-3}) has to be satisfied. Considering inequalities
(\ref{eq: mac-1}) and (\ref{eq: mac-3}), it is sufficient to show
that $t_{3}C_{02}\leq \bar{t}_{3}C_{23}$. Using the values obtained in Eq. (\ref{eqnarr: RMAC}), this is equivalent to prove:
\begin{equation*}
    C_{02}\big(C_{01}(C_{\text{MAC}}-C_{13})+C_{13}C_{23}\big)\leq C_{23}\big(\Delta+C_{13}(C_{\text{MAC}}-C_{13}+C_{02})\big).
\end{equation*}
 By re-ordering the terms and using the definition of $\Delta$, the above inequality can be alternatively written as:
\begin{equation*}
    C_{\text{MAC}} \Delta\leq (C_{13}+C_{23})\Delta,
\end{equation*}
 which is true since $\Delta\!>\!0$, and $C_{\text{MAC}}=\mathcal{C}(g_{13}+g_{23})$.

For $\Gamma'\!>\!0$, inequalities (\ref{eq: mac-1}), (\ref{eq:
mac-3}), and (\ref{eq: mac-4}) are set to equality. In this case,
the time intervals and the achievable rate become:
\begin{IEEEeqnarray}{rl}
t_{2}&=\frac{C_{02}(C_{\text{MAC}}-C_{23})+C_{13}C_{23}}{(C_{02}+C_{23})(C_{\text{MAC}}-C_{23}+C_{01})},\nonumber\\
t_{3}&=\frac{C_{23}}{C_{02}+C_{23}},\nonumber\\
t_{4}&=\frac{\Delta}{(C_{02}+C_{23})(C_{\text{MAC}}-C_{23}+C_{01})},\nonumber\\
R_{\textrm{MDF-MAC}}^{2}&=\frac{C_{02}(C_{01}+C_{23})}{C_{02}+C_{23}}-\frac{ C_{01}\Delta}{(C_{02}+C_{23})(C_{\text{MAC}}-C_{23}+C_{01})}.
\end{IEEEeqnarray}
\subsection{Proof of Lemma \ref{lemma: C123-CMAC}}\label{appendix: lemma6}
\begin{IEEEeqnarray*}{rl}
C_{123}-C_{\text{MAC}}&=\frac{1}{2}\log\Big(\frac{1+(\sqrt{g_{13}}+\sqrt{g_{23}})^{2}}{1+g_{13}+g_{23}}\Big)\\
{}&=\frac{1}{2}\log\Big(1+\frac{2\sqrt{g_{13}\ g_{23}}}{1+g_{13}+g_{23}}\Big)\\
{}&\leq \frac{1}{2}\log\Big(1+\frac{g_{13}+g_{23}}{1+g_{13}+g_{23}}\Big)\\
{}&\leq \frac{1}{2}.
\end{IEEEeqnarray*}
\subsection{Proof of Lemma \ref{lemma: DiffC012}}\label{appendix2}
\begin{IEEEeqnarray*}{ll}
C_{012}-\max\{C_{01},C_{02}\}&=\frac{1}{2}\log\left(\frac{1+g_{01}+g_{02}}{1+\max\{g_{01},g_{02}\}}\right)\nonumber\\
{}&=\frac{1}{2}\log\left(1+\frac{\min\{g_{01},g_{02}\}}{1+\max\{g_{01},g_{02}\}}\right)\nonumber\\
{}&\leq\frac{1}{2}\log\left(1+\frac{\max\{g_{01},g_{02}\}}{1+\max\{g_{01},g_{02}\}}\right)\nonumber\\
{}&\leq \frac{1}{2},
 \end{IEEEeqnarray*}
\begin{IEEEeqnarray*}{ll}
C_{123}-\max\{C_{13},C_{23}\}&=\frac{1}{2}\log\left(\frac{1+(\sqrt{g_{13}}+\sqrt{g_{23}})^{2}}{1+\max\{g_{13},g_{23}\}}\right)\nonumber\\
{}&=\frac{1}{2}\log\left(1+\frac{\min\{g_{13},g_{23}\}+2\sqrt{g_{13}g_{23}}}{1+\max\{g_{13},g_{23}\}}\right)\nonumber\\
{}&\leq\frac{1}{2}\log\left(1+\frac{3\sqrt{g_{13}g_{23}}}{1+\max\{g_{13},g_{23}\}}\right)\nonumber\\
{}&\leq \frac{1}{2}\log\left(1+\frac{3\sqrt{g_{13}g_{23}}}{1+\sqrt{g_{13}g_{23}}}\right)\nonumber\\
{}&\leq 1.
 \end{IEEEeqnarray*}
 \subsection{Proof of Lemma \ref{lemma: kappa5}}\label{appendix: kappa5}
 In this region, $C_{01}\leq 1$ and $C_{01}\leq C_{02}$, therefore, $0 \leq C_{13}C_{23}(C_{02}-C_{01})(1-C_{01})$. It is easy to verify that the following inequality is valid:
 \begin{equation}
    2C_{13}C_{23}\big(C_{01}(C_{02}-C_{01})+0.5(C_{01}+C_{23})\big)\leq (C_{01}+C_{13})(C_{02}+C_{23})\big(C_{23}+.5+C_{01}(C_{02}-C_{01})\big).
    \end{equation}
     Replacing $C_{13}C_{23}$ by the smaller quantity $(C_{13}C_{23}-C_{01}C_{02})$ in the LHS of the above inequality results in:
 \begin{equation}
    2(C_{13}C_{23}-C_{01}C_{02})\big(C_{01}(C_{02}-C_{01})+0.5(C_{01}+C_{23})\big)\leq (C_{01}+C_{13})(C_{02}+C_{23})\big(C_{23}+.5+C_{01}(C_{02}-C_{01})\big).
    \end{equation}
Since $C_{01}\leq 1$ in the RHS, $C_{01}(C_{02}-C_{01})$ can be
substituted by the larger term $(C_{02}-C_{01})$. Hence, the
following inequality is true:
 \begin{equation}
    -2\Delta\big(C_{01}(C_{02}-C_{01})+0.5(C_{01}+C_{23})\big)\leq (C_{01}+C_{13})(C_{02}+C_{23})\big(C_{23}+.5+(C_{02}-C_{01})\big).
    \end{equation}
Rearranging the terms leads to:
\begin{equation}
    \frac{-\Delta}{C_{01}+C_{13}}
\Big(\frac{C_{01}+C_{23}}{C_{02}+C_{23}}-\frac{C_{23}}{C_{02}+0.5-C_{01}+C_{23}}\Big)\leq \frac{1}{2}.
\end{equation}
The gap can be further increased by replacing $C_{02}+0.5$ with the smaller term $C_{012}$ according to Lemma \ref{lemma: DiffC012}. Therefore:
\begin{equation}
    \frac{-\Delta}{C_{01}+C_{13}}
\Big(\frac{C_{01}+C_{23}}{C_{02}+C_{23}}-\frac{C_{23}}{C_{012}-C_{01}+C_{23}}\Big)\leq \frac{1}{2},
\end{equation}
which completes the proof.
\subsection{Proof of Lemma \ref{lemma: kappa7}}\label{appendix: kappa7}
\begin{IEEEeqnarray*}{ll}
\kappa_{7}&=\frac{\Delta}{C_{01}+C_{13}}
\Big(\frac{C_{02}+C_{13}}{C_{02}+C_{23}}-\frac{C_{02}}{C_{123}-C_{13}+C_{02}}\Big)\nonumber\\
{}&\stackrel{(a)}{\leq}\frac{\Delta}{C_{01}+C_{13}}\times\frac{C_{13}}{C_{02}+C_{23}}+\delta\nonumber\\
{}&\stackrel{(b)}{\leq}\frac{\Delta}{(C_{01}+C_{13})(C_{02}+C_{23})}+\delta\nonumber\\
{}&\stackrel{(c)}{\leq}\frac{C_{01}}{C_{01}+C_{13}}\times \frac{C_{02}}{C_{02}+C_{23}}+\delta\nonumber\\
{}&\leq1+\delta.
\end{IEEEeqnarray*}
As $C_{123}\leq C_{13}+C_{23}$ in this region, $C_{123}-C_{13}$ is
replaced by the larger quantity $C_{23}$ to obtain $(a)$. $(b)$ is
valid since $C_{13}\leq 1$ for this scenario. In $(c)$, $\Delta$ is
substituted by the larger term $C_{01}C_{02}$.
\section{Gap Analysis Summary}\label{appendix Table}
The results related to gap analysis are compactly shown in Table I.
For each region specified by some conditions on the link capacities,
the corresponding symbols for the upper bound, the achievable rate,
and  the gap, (\emph{i.e.}, the difference between the upper bound
and the achievable rate) are shown \footnote{The characterizing
equation for each symbol used in the table is given in the body of
the paper.}. In addition, an upper bound on the value of the gap is
given. For instance, for the region specified by $\Delta\leq0,
\Gamma\leq0$, and $C_{02}\leq C_{01}$ conditions, the upper bound,
the achievable rate, and the gap are respectively represented by
$R_{\text{up}}^{1}$, $R_{\text{MDF}}^{1}$, and $\kappa_{1}$. Using
the achievable scheme that leads to $R_{\text{MDF}}^{1}$, the gap
from the upper bound $R_{\text{up}}^{1}$ is less than
$\frac{1}{2}+\delta$. Our results, summarized in Table I, indicate
that sending \emph{independent} information
 during each mode together with the decode-and-forward scheme are sufficient to operate close to the capacity
 of the channel.
\begin{table}[!tbh]
\caption{Summary of the Results: Gap Analysis for Different Regions}
\label{tbl: SMALL-GAP}
\centering
\begin{tabular}[c]{|c|c|c|c|c|c|c|c|c|}\hline
\multicolumn{4}{|c|}{\multirow{2}{*}{Region}} & \multirow{2}{*}{Achievable Rate} & \multirow{2}{*}{Gap} & Upper Bound & Upper Bound\\
\multicolumn{4}{|c|}{} & & & on the Gap & on the Capacity \\ \hline
\multirow{3}{*}{$\Delta\leq0$} & \multirow{3}{*}{$\Gamma\leq0$} & \multicolumn{2}{c|}{$C_{02}\leq C_{01}$} & $R_{\text{MDF}}^{1}$ & $\kappa_{1}$ & \multirow{3}{*}{$\frac{1}{2}+\delta$} &\multirow{3}{*}{$R_{\text{up}}^{1}$}  \\ \cline{3-6}
& & \multirow{2}{*}{$C_{02}\geq C_{01}$} & $C_{01}\leq 1$ &$R_{\text{MDF}}^{2}$ & $\kappa_{5}$ & & \\ \cline{4-6}
& & & $C_{01}\geq 1$ & $R_{\text{MDF-BC}}^{1}$ & $\kappa_{\text{MDF-BC}}^{1}$ & &\\ \hline \hline
\multirow{3}{*}{$\Delta\leq0$} & \multirow{3}{*}{$\Gamma>0$} & \multicolumn{2}{c|}{$C_{01}\leq C_{02}$} & $R_{\text{MDF}}^{2}$ & $\kappa_{2}$ & \multirow{3}{*}{$\frac{1}{2}+\delta$} & \multirow{3}{*}{$R_{\text{up}}^{2}$} \\
\cline{3-6}
& & \multirow{2}{*}{$C_{01}\geq C_{02}$} & $C_{02}\leq 1$ &$R_{\text{MDF}}^{1}$ & $\kappa_{6}$ & & \\ \cline{4-6}
& & & $C_{02}\geq 1$ & $R_{\text{MDF-BC}}^{2}$ & $\kappa_{\text{MDF-BC}}^{2}$ & &\\ \hline \hline
\multirow{6}{*}{$\Delta>0$} & \multirow{6}{*}{$\Gamma'\leq0$} & \multicolumn{2}{c|}{\multirow{2}{*}{$C_{23}\leq C_{13}$}} & $R_{\text{MDF}}^{3}$ & $\kappa_{3}$ & $1+\delta$ & \multirow{6}{*}{$R_{\text{up}}^{3}$} \\ \cline{5-7}
& & \multicolumn{2}{c|}{} & $R_{\text{MDF-MAC}}^{1}$ & $\kappa_{\text{MDF-MAC}}^{1}$ & $\frac{1}{2}+\delta$ &\\ \cline{3-7}
& & \multirow{4}{*}{$C_{23}\geq C_{13}$} & \multirow{2}{*}{$C_{13}\leq 1$, $C_{123}\leq C_{13}+C_{23}$} &$R_{\text{MDF}}^{4}$ & $\kappa_{7}$ & $1$ &\\ \cline{5-7}
& & & & \multirow{3}{*}{$R_{\text{MDF-MAC}}^{1}$} & \multirow{3}{*}{$\kappa_{\text{MDF-MAC}}^{1}$} & $\frac{1}{2}$ &\\ \cline{4-4} \cline{7-7}
& & & $C_{13}\leq 1$, $C_{123}\geq C_{13}+C_{23}$ & & & \multirow{2}{*}{$\frac{1}{2}+\delta$} & \\ \cline{4-4}
& & & $C_{13}\geq 1$ & & & &\\ \hline \hline
\multirow{6}{*}{$\Delta>0$} & \multirow{6}{*}{$\Gamma'>0$} & \multicolumn{2}{c|}{\multirow{2}{*}{$C_{13}\leq C_{23}$}} & $R_{\text{MDF}}^{4}$ & $\kappa_{4}$ & $1+\delta$ & \multirow{6}{*}{$R_{\text{up}}^{4}$} \\ \cline{5-7}
& & \multicolumn{2}{c|}{} & $R_{\text{MDF-MAC}}^{2}$ & $\kappa_{\text{MDF-MAC}}^{2}$ & $\frac{1}{2}+\delta$ &\\ \cline{3-7}
& & \multirow{4}{*}{$C_{13}\geq C_{23}$} & \multirow{2}{*}{$C_{23}\leq 1$, $C_{123}\leq C_{13}+C_{23}$} &$R_{\text{MDF}}^{3}$ & $\kappa_{8}$ & $1$ &\\ \cline{5-7}
& & & & \multirow{3}{*}{$R_{\text{MDF-MAC}}^{2}$} & \multirow{3}{*}{$\kappa_{\text{MDF-MAC}}^{2}$} & $\frac{1}{2}$ &\\ \cline{4-4} \cline{7-7}
& & & $C_{23}\leq 1$, $C_{123}\geq C_{13}+C_{23}$ & &  & \multirow{2}{*}{$\frac{1}{2}+\delta$} & \\ \cline{4-4}
& & & $C_{23}\geq 1$ & & & &\\ \hline
\end{tabular}
\end{table}

\end{document}